\makeatletter\def\input@path{
{Figures/}
}\makeatother%
\documentclass[journal=jcisd8,manuscript=article,keywords=true]{achemso}

\usepackage[utf8]{inputenc}
\usepackage[T1]{fontenc}

\usepackage{geometry}
\usepackage{setspace}

\usepackage{graphicx}
\usepackage{float}
\newfloat{scheme}{htbp}{los}
\floatname{scheme}{Scheme}
\floatname{chart}{Chart}
\newfloat{graph}{htbp}{loh}

\usepackage{chemformula,chemfig}
\setchemfig{atom style={scale=0.4}, double bond sep = 4pt}
\usepackage[version=4,arrows=pgf-filled,textfontname=sffamily,
mathfontname=mathsf]{mhchem}
\usepackage{microtype}  
\usepackage[detect-all=true,group-minimum-digits=4]{siunitx}
\usepackage{tabularx, longtable}%
\usepackage{booktabs, multirow,xcolor,cleveref,url}

\usepackage{amssymb}
\usepackage{seqsplit}   
\usepackage{xr-hyper}  
\usepackage{threeparttable}

\newcommand{\smiles}[1]{\parbox{6.5cm}{\seqsplit{\ttfamily#1}}}
\newcommand{\deltaest}{$\Delta E_{\text{ST}}$}

\DeclareSIUnit\calorie{cal}
\DeclareSIUnit\kcal{\kilo\calorie}
\DeclareSIUnit\kcal{Kcal}
\DeclareSIUnit\atomicunit{a.u.}

\author{Jean-Pierre Tchapet Njafa}
\email{jean-pierre.tchapet@facsciences-uy1.cm}
\affiliation{Department of Physics, Faculty of Science, University of Yaounde I, Yaounde 812, Yaounde, Cameroon}

\author{Elvira Vanelle Kameni Tcheuffa}
\affiliation{Department of Physics, Faculty of Science, University of Yaounde I, Yaounde 812, Cameroon}

\author{Aissatou Maghame Foumkpou}
\affiliation{Department of Physics, Faculty of Science, University of Yaounde I, Yaounde 812, Cameroon}

\author{Serge Guy Nana Engo}
\affiliation{Department of Physics, Faculty of Science, University of Yaounde I, Yaounde 812, Yaounde, Cameroon}

\title{Data-Driven Design Guidelines for TADF Emitters from a High-Throughput Screening of
\num{747} Molecules}


\begin{document}

\maketitle

\begin{abstract}
TADF emitter performance depends on both thermodynamic and kinetic factors. We analyze 747 experimentally known TADF molecules computationally to extract 
quantitative design guidelines. Using a validated xTB-based workflow, we examine how architecture, geometry, and electronic structure affect photophysical 
properties. Among architectures, D-A-D frameworks achieve the smallest \deltaest. A favorable torsional angle of \qtyrange{50}{90}{\degree} balances small 
\deltaest{} with the spin--orbit coupling needed for reverse intersystem crossing. Clustering separates high-performance candidates and highlights 
multi-resonance emitters for blue emission. From these results, we identify 127 candidates with predicted \deltaest{} $<\qty{0.1}{\electronvolt}$ and oscillator 
strength $f > 0.1$. These HTVS-derived design guidelines and candidates can guide future TADF emitter development.
\end{abstract}


\section{INTRODUCTION}\label{sec:introduction}

Organic Light-Emitting Diodes (OLEDs) based on Thermally Activated Delayed Fluorescence (TADF) have defined the third generation of electroluminescent 
materials, offering a viable path to achieving near-unity internal quantum efficiency (IQE) without resorting to the expensive and scarce noble metals found in 
phosphorescent emitters \cite{Uoyama2012, Adachi2014TADF, Liu2018}. The TADF mechanism hinges on the efficient harvesting of non-emissive triplet excitons 
($T_1$) by up-converting them to the emissive singlet state ($S_1$) through Reverse Intersystem Crossing (RISC). This process requires satisfying both a 
thermodynamic condition—a minimized singlet-triplet energy gap (\deltaest $\lesssim \qty{0.2}{\electronvolt}$)—and a kinetic condition: a sufficiently high rate 
of RISC ($k_{\text{RISC}}$) \cite{Dias2017, Chen2018}.

The rational design of efficient TADF emitters faces an inherent trade-off that complicates optimization. In conventional Donor-Acceptor (D-A) architectures, 
achieving a small \deltaest necessitates a significant spatial separation of the frontier molecular orbitals to minimize the electron exchange energy 
\cite{Moral2015}. However, this separation inherently reduces the orbital overlap, which not only suppresses the oscillator strength ($f$) and thus the 
radiative decay rate, but can also hinder the spin-orbit coupling (SOC) necessary for efficient RISC \cite{Penfold2015, Olivier2017}. Optimizing both properties 
simultaneously demands exploration across diverse molecular scaffolds \cite{Gibson2016}.

Computational methods can accelerate this search, but they introduce a cost-accuracy trade-off. High-fidelity methods are computationally prohibitive for
screening the thousands of potential candidates required for data-driven discovery
\cite{Sun2015, Kang2024, Olivier2017a}. This cost barrier has confined most studies to small molecular series. Universal design guidelines are therefore still 
lacking.

Here, we employ a high-throughput virtual screening (HTVS) protocol based on accelerated semiempirical methods. We validated this approach on a data set of 
\num{747} experimentally known TADF emitters in our companion paper (hereafter, Article 1 \cite{TchapetArxivA1-2025}), which established its reliability for 
large-scale molecular ranking and trend analysis.

In this article, we leverage our validated HTVS protocol to move beyond benchmarking and perform a large-scale analysis of structure-property relationships 
governing TADF performance. By analyzing the entire 747-molecule data set, we aim to establish quantitative design guidelines from the interplay of molecular 
structure, conformation, and electronic properties. We aim to establish an architecture ranking, derive structure-property rules for both \deltaest{} and 
$k_{\text{RISC}}$, classify molecules by clustering, and propose synthetic targets with favorable predicted properties. From this data set, we extract practical 
design guidelines. These can inform the synthesis of improved TADF materials.

This paper is structured as follows: Section \ref{sec:methods} details our
methodological framework. The main results are presented in the subsequent sections,
including an analysis of the architectural landscape, a classification of molecular
families, and a deep dive into the MR-TADF design paradigm. Finally,
Section \ref{sec:conclusions} summarizes our conclusions and proposes future directions
for TADF emitter design.

\section{METHODOLOGICAL FRAMEWORK FOR LARGE-SCALE ANALYSIS}\label{sec:methods}

All structure-property relationships presented in this work were derived from a large-scale data set of \num{747} experimentally known TADF emitters, processed 
using a consistent and validated high-throughput virtual screening (HTVS) protocol. The full details, validation, and limitations of this computational 
framework are established in our preceding work (Article 1 \cite{TchapetArxivA1-2025}). Here, we provide a concise summary of the essential methodological 
components and their justifications.

\subsection{Data Set and Molecular Structures}

The data set comprises a diverse set of \num{747} TADF molecules curated from the literature \cite{Huang_2024}. To facilitate systematic analysis, molecules 
were classified into distinct architectural families (e.g., D-A, D-A-D, Multiresonance) using automated SMARTS pattern matching, followed by manual 
verification.

Initial 3D structures were generated from SMILES strings, and a rigorous conformational search was performed for each molecule using the CREST program coupled 
with the GFN2-xTB semiempirical Hamiltonian \cite{pracht2020automated, Bannwarth2019}. The lowest-energy conformer was then fully optimized at the GFN2-xTB 
level to obtain the final ground-state ($S_0$) geometry. This tight-binding method is parametrized for molecular structures and noncovalent interactions, which 
is appropriate for the flexible systems in our data set \cite{bannwarth2021extended}.

\subsection{Methodological Approximations and Justifications}

\subsubsection{Hybrid Protocol and the Vertical Approximation Compromise}

The HTVS protocol relies on a hybrid computational strategy: ground-state ($S_0$) geometries are optimized using GFN2-xTB \cite{Bannwarth2019, Ehlert2021}, 
while excited-state properties ($S_1$, $T_1$) are calculated using the simplified methods, sTDA-xTB/sTD-DFT-xTB \cite{Grimme2016, Wergifosse2024}. This 
methodology implicitly employs the vertical approximation.

While GFN2-xTB is highly effective for determining ground-state structures, we acknowledge
that applying the $S_0$ geometry to compute vertical gaps neglects excited-state geometry
relaxation ($S_1$/$T_1$), which is important for accurately describing adiabatic
properties, particularly the \deltaest~ in TADF systems \cite{Mewes2018, Olivier2017}.
This compromise reduces computational cost by over $\qty{99}{\percent}$ relative to TD-DFT
\cite{TchapetArxivA1-2025}. To validate this approximation, we performed TD-DFT excited-state geometry optimizations for 6 representative molecules (see 
Supporting Information). The results show moderate correlation between xTB adiabatic estimates and TD-DFT adiabatic values ($R^2 = 0.40$--0.45, Spearman $\rho = 
0.54$--0.66), confirming that molecular rankings are preserved despite systematic differences. The resulting systematic errors, characterized in Article 1
\cite{TchapetArxivA1-2025}, do not affect the relative molecular rankings needed for
screening and trend analysis. Environmental effects were included via a generalized
Born (GB) model with solvent
accessible surface (SASA) (GBSA) implicit solvation model for toluene ($\varepsilon =
2.38$) \cite{Ehlert2021}.

\subsubsection{Solvation Model Selection and Trade-offs}

The treatment of solvation effects in excited-state calculations presents a fundamental trade-off between computational efficiency and physical accuracy, 
particularly for charge-transfer (CT) states characteristic of TADF emitters. This study employed the ground-state-based GBSA implicit solvation model for 
toluene, which warrants detailed discussion of its limitations and the rationale for this choice in high-throughput screening applications.

\paragraph{Ground-state Vs. State-specific Solvation Models}

Traditional implicit solvation models like GBSA utilize the ground-state electron density to compute solvation energies for all electronic 
states.\cite{Marenich2009} While computationally efficient, this approximation becomes problematic for CT states where significant charge redistribution occurs 
upon excitation.\cite{Chibani2012} The solvent response to electronic excitation can differ substantially from ground-state solvation, as the reorganized 
electron density in CT states creates different electrostatic interactions with the surrounding medium.\cite{Improta2006}

State-specific solvation models, such as the state-specific polarizable continuum model (SS-PCM), address this limitation by computing separate solvation 
energies for each electronic state using the corresponding electron density.\cite{Improta2006,Tomasi2005} For TADF emitters, where the $S_1$ and $T_1$ states 
often exhibit pronounced CT character with substantial charge separation between donor and acceptor moieties, SS-PCM can provide significantly more accurate 
solvation energies.\cite{Moral2015}

\paragraph{Implications for Tadf Properties}

The choice of solvation model has direct implications for predicted TADF properties. Solvent polarity modulates the singlet-triplet energy gap (\deltaest) 
through differential solvation of the $S_1$ and $T_1$ states.\cite{ArticleRef1} CT states, with their spatially separated charges, typically exhibit stronger 
solvation effects than locally excited states. Consequently, ground-state-based models may systematically under- or overestimate \deltaest~ values, 
particularly in polar solvents.

Recent computational studies of TADF emitters using state-specific approaches have demonstrated that solvent effects can alter \deltaest~ by 0.1-0.3 eV 
compared to gas-phase calculations, with the magnitude depending on the degree of charge separation and solvent polarity.\cite{Moral2015} Ground-state-based 
models may capture the overall trend but with reduced quantitative accuracy.

\paragraph{Computational Cost Considerations}

The primary limitation of state-specific solvation models is computational cost. SS-PCM requires separate self-consistent field calculations for each electronic 
state, typically increasing computational time by a factor of 2-3 compared to ground-state models.\cite{Tomasi2005} For high-throughput screening of 747 
molecules with multiple conformers, this translates to thousands of additional calculations, making SS-PCM computationally prohibitive within reasonable time 
and resource constraints.

\paragraph{Validation of Gbsa for Screening Applications}

Despite its limitations, GBSA has been extensively validated for screening applications in drug discovery and materials science.\cite{Genheden2015} While 
absolute solvation energies may contain systematic errors, the relative ordering of molecules and structure-property trends are often preserved. This makes GBSA 
suitable for the primary objective of this study: identifying structural patterns and relative rankings rather than predicting absolute TADF efficiencies.

The key assumption underlying our approach is that the systematic errors introduced by the ground-state approximation affect all molecules similarly, preserving 
the relative trends essential for deriving design guidelines. This assumption is supported by the observation that GBSA models provide reliable relative free 
energies for structurally similar compounds.\cite{Genheden2015}

\paragraph{Limitations and Future Directions}

We acknowledge several limitations of the GBSA approach for TADF emitters:

\begin{enumerate}
\item \textbf{Quantitative accuracy}: Absolute \deltaest~ values may deviate from experimental measurements due to inadequate treatment of CT state solvation.
\item \textbf{Solvent dependence}: The model cannot capture how different solvents might alter the relative rankings of molecules.
\item \textbf{Strong CT states}: Molecules with highly polarized CT states may be most affected by the ground-state approximation.
\end{enumerate}

Future studies should validate these findings using state-specific solvation models for representative subsets of molecules, particularly those identified as 
promising TADF candidates. Additionally, experimental validation in different solvents would help assess the transferability of the derived design guidelines.

\paragraph{Justification for Current Approach}

Given the screening nature of this study and the computational constraints, GBSA represents an appropriate balance between accuracy and efficiency. The model 
enables the systematic analysis of 747 molecules while maintaining sufficient accuracy to identify meaningful structure-property relationships. The derived 
guidelines should be viewed as screening criteria for identifying promising candidates, with the understanding that quantitative predictions require more 
sophisticated solvation treatments.

\subsubsection{Specific Limitations for Multi-resonance Emitters}

MR-TADF emitters, characterized by their extended $\pi$-conjugation and multiple resonance structures involving boron-nitrogen doping, are known to exhibit 
multi-reference character in their electronic structure.\cite{Hatakeyama2016,Kondo2019} While single-reference methods like sTD-DFT-xTB may not fully capture 
the quantitative energetics of these systems, previous validation studies have demonstrated that qualitative trends in \deltaest~ and orbital properties remain 
consistent between semiempirical, TD-DFT, and post-Hartree-Fock methods.\cite{Pershin2019,Olivier2022}

Our data set includes several MR-TADF molecules (e.g., 24PQ-Cz, phenazasiline), which we validated against full TD-DFT calculations using the CAM-B3LYP 
functional with the def2-TZVP basis set (see Supporting Information Table S6). While absolute \deltaest~ values show systematic deviations (MAE = 
\qty{1.28}{\electronvolt}), the relative ranking of molecules is preserved (Table S6), confirming that our screening protocol is appropriate for identifying and 
prioritizing promising MR-TADF candidates. This is consistent with the established understanding that high-throughput virtual screening methods capture trends 
rather than quantitative predictions,\cite{Gomez-Bombarelli2016} and we recommend post-hoc validation with multi-reference methods (e.g., STEOM-DLPNO-CCSD, 
CASPT2) for quantitative predictions of specific candidates before experimental synthesis.

\paragraph{Spin-orbit Coupling Calculations}

SOC matrix elements were calculated using ORCA 6.1.0\cite{Neese2022} with the TD-DFT/TDA approach and the SOMF(1X) spin-orbit operator.\cite{Neese2005} 
Calculations employed the $\omega$B97X-D4 functional\cite{Najibi2018} with the def2-TZVP basis set\cite{Weigend2005} and included solvent effects (toluene) via 
the CPCM model.\cite{Barone1998} The SOC matrix was constructed between the lowest five singlet and five triplet excited states, and T$_1$--S$_1$ coupling 
elements were extracted for analysis.

\subsection{Derivation of Structure-property Descriptors and Theoretical Framework}

To translate the large computational data set into actionable design rules, donor-acceptor (D-A) torsional angles were automatically extracted from all 
optimized geometries. To quantify the degree of charge-transfer and its relation to geometry, we calculated the HOMO-LUMO overlap integral ($S'_{\text{HL}}$) 
and the distance between the centroids of the hole and electron densities ($\Delta r_{\text{CT}}$) using the Multiwfn package \cite{lu2024multiwfn, 
Etienne2014}. Principal Component Analysis (PCA) was performed on the standardized matrix of all descriptors to identify the underlying variables governing TADF 
performance.

This analysis is grounded in the theoretical framework of TADF efficiency, which depends
not only on a small \deltaest~ (thermodynamics) but also on a high rate of reverse
intersystem crossing ($k_{\text{RISC}}$), a kinetic parameter governed by spin-orbit
coupling (SOC). According to El-Sayed's rules, SOC is maximized when the interconverting
singlet ($S_1$) and triplet ($T_1$) states have different orbital characters (e.g.,
charge-transfer vs. local-excitation) \cite{El-Sayed-Rules-1968,
Li_2022_10_1016_j_orgel_2022_106645}. Our chosen descriptors—particularly the D-A
torsional angle—are direct proxies for tuning this CT/LE character balance. While our
protocol does not compute SOC explicitly due to its prohibitive cost, analyzing these
structural and electronic descriptors provides a robust, qualitative assessment of the
kinetic factors that, alongside \deltaest, govern overall TADF performance.

\section{RESULTS AND DISCUSSION: FROM STATISTICAL TRENDS TO RATIONAL DESIGN PRINCIPLES}
\label{sec:Res_Disc}

This section leverages the large-scale data set of \num{747} TADF emitters, processed with the validated HTVS protocol detailed in our companion paper (Article 
1 \cite{TchapetArxivA1-2025}), to extract quantitative, statistically robust structure-property relationships. Our analysis transitions from broad statistical 
trends to specific, actionable molecular design rules, with a focus on providing a deep mechanistic understanding that connects molecular structure to the key 
photophysical parameters governing TADF efficiency.

\subsection{Low-dimensional Nature of the Tadf Design Space}

\begin{figure}[!ht]
\centering
\includegraphics[width=0.9\textwidth]{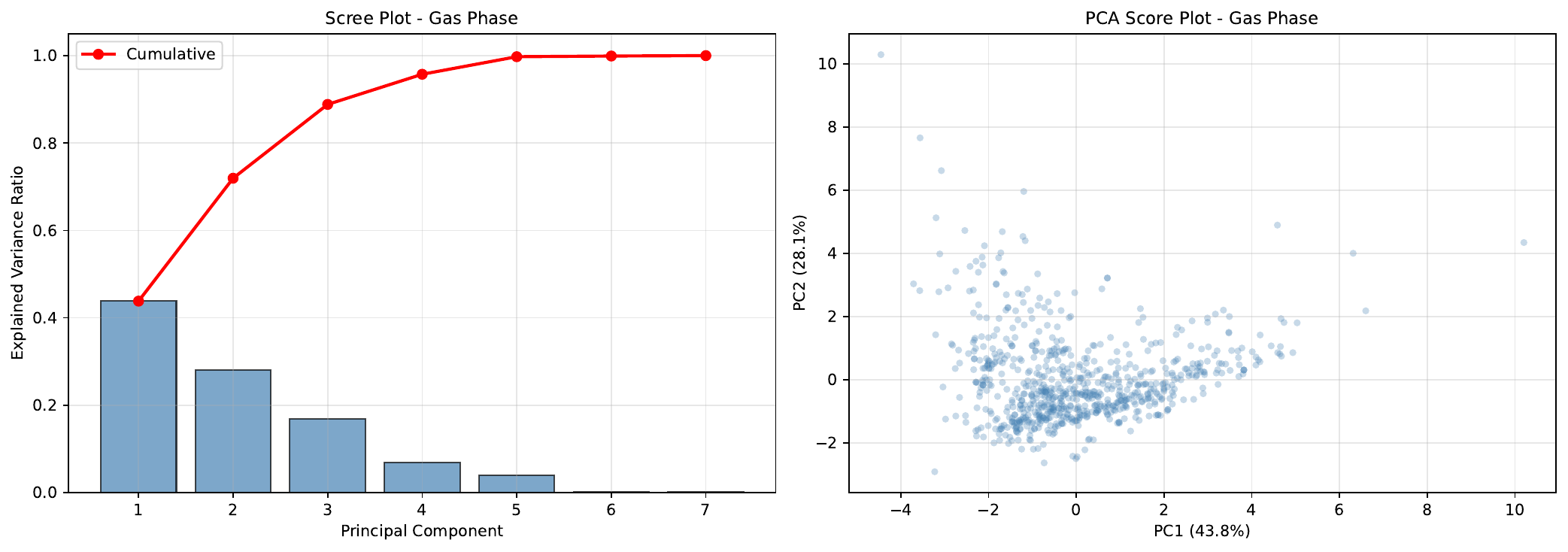}
\caption{Principal component analysis score plots for the \num{747} TADF emitters in the gas phase. The tight clustering of data along the first two principal 
components, which collectively capture \qty{71.9}{\percent} of the variance, illustrates the low intrinsic dimensionality of the TADF property space. This 
finding suggests that the complex design challenge can be rationalized by optimizing a few key orthogonal properties.}
\label{fig:pca_analysis}
\end{figure}

To distill the complex, high-dimensional relationships between molecular structure and photophysical properties, we first performed a principal component 
analysis (PCA) on the full data set. Despite the chemical diversity, the property space proves to be low-dimensional. As shown in \Cref{fig:pca_analysis}, the 
first three principal components (PCs) capture a cumulative variance of \qty{88.8}{\percent}, indicating that the majority of performance variation is governed 
by a small number of latent factors. These components correspond to intuitive physical properties: PC1 (\qty{43.8}{\percent}) relates to the fundamental 
excitation energy (HOMO-LUMO gap and $\lambda_{\text{PL}}$), PC2 (\qty{28.1}{\percent}) describes the degree of charge-transfer character (related to 
\deltaest), and PC3 (\qty{16.9}{\percent}) is associated with the electronic coupling and oscillator strength. This low dimensionality suggests that predictive 
modeling and rational design are feasible for TADF systems.

\subsection{Architectural Hierarchy Defines Tadf Potential}

\begin{figure}[!htbp]
\centering
\includegraphics[width=0.9\textwidth]{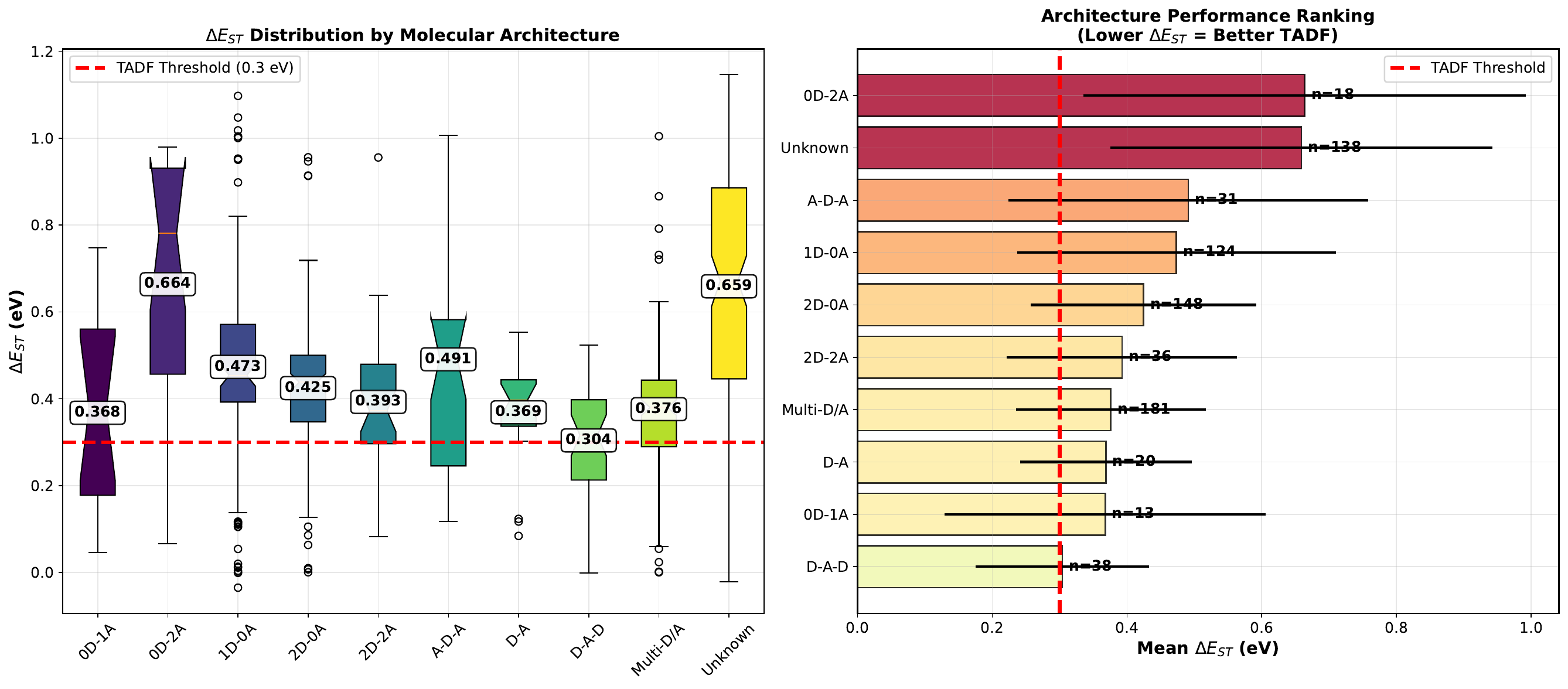}
\caption{TADF performance stratified by molecular architecture. The box plots show the distribution of the singlet-triplet energy gap (\deltaest) for each 
class. D-A-D architectures demonstrate statistically superior characteristics (lower median and narrower distribution of \deltaest) compared to other systems, 
confirming the efficacy of this design principle for consistently achieving small energy gaps.}
\label{fig:performance_hierarchy}
\end{figure}

To understand the physical origin of the principal components, we systematically
classified the molecules and evaluated their performance by architectural class. This
analysis reveals a clear performance hierarchy (\Cref{fig:performance_hierarchy}).
Donor-Acceptor-Donor (D-A-D) systems are statistically superior, exhibiting a mean
\deltaest~ of \qty{0.30}{\electronvolt}, significantly lower than that of simple D-A
(\qty{0.37}{\electronvolt}) or multi-D/A systems (\qty{0.38}{\electronvolt}). This
confirms that architectures promoting symmetric charge delocalization are highly effective
at minimizing the exchange energy, a cornerstone of TADF theory.

\subsection{Data-driven Clustering Reveals Distinct Molecular Families}

\begin{figure}[!ht]
 \begin{minipage}{0.49\textwidth}
 \centering
  \includegraphics[width=\textwidth]{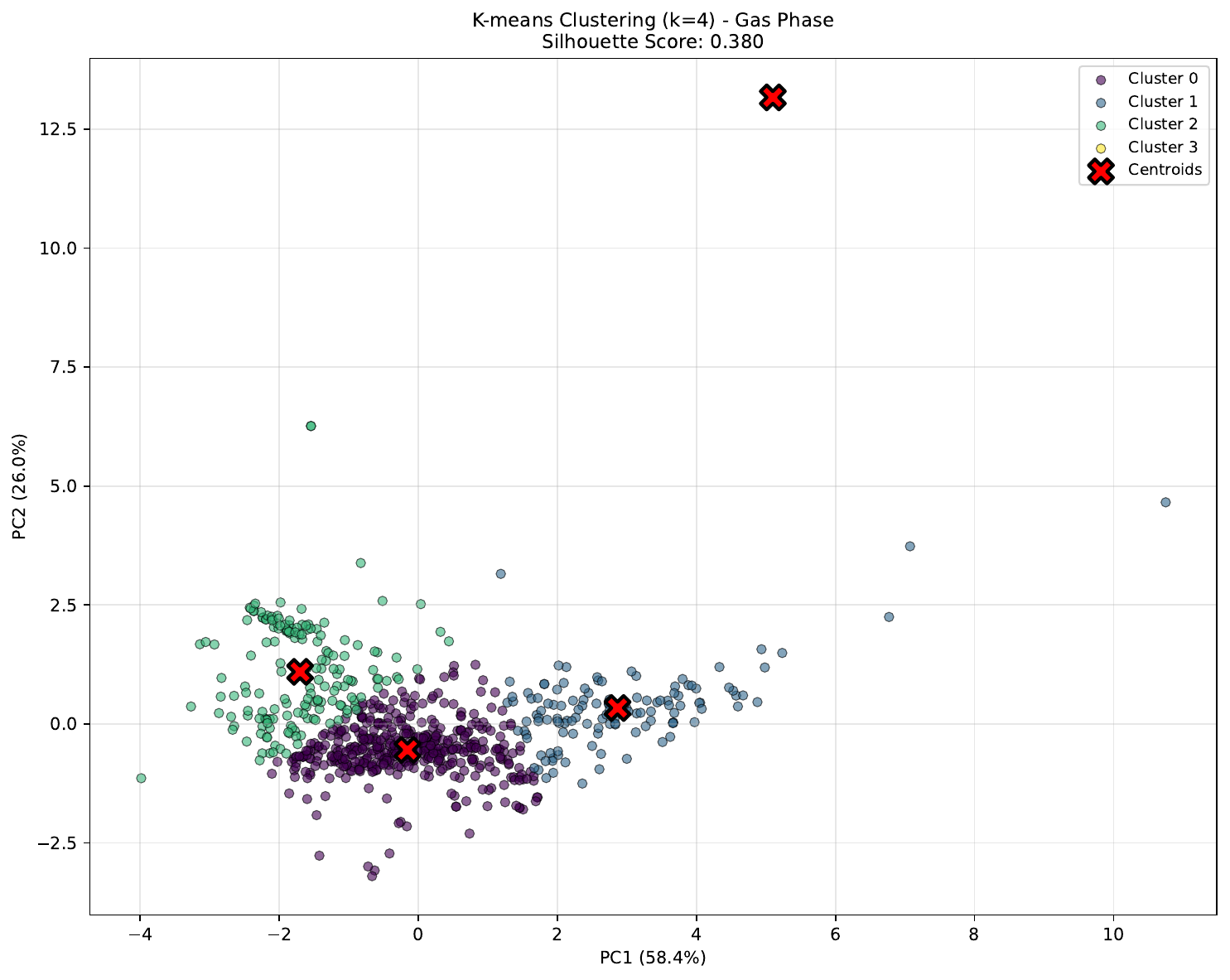}\\(a)
 \end{minipage}\hfill
 \begin{minipage}{0.49\textwidth}
 \centering
  \includegraphics[width=\textwidth]{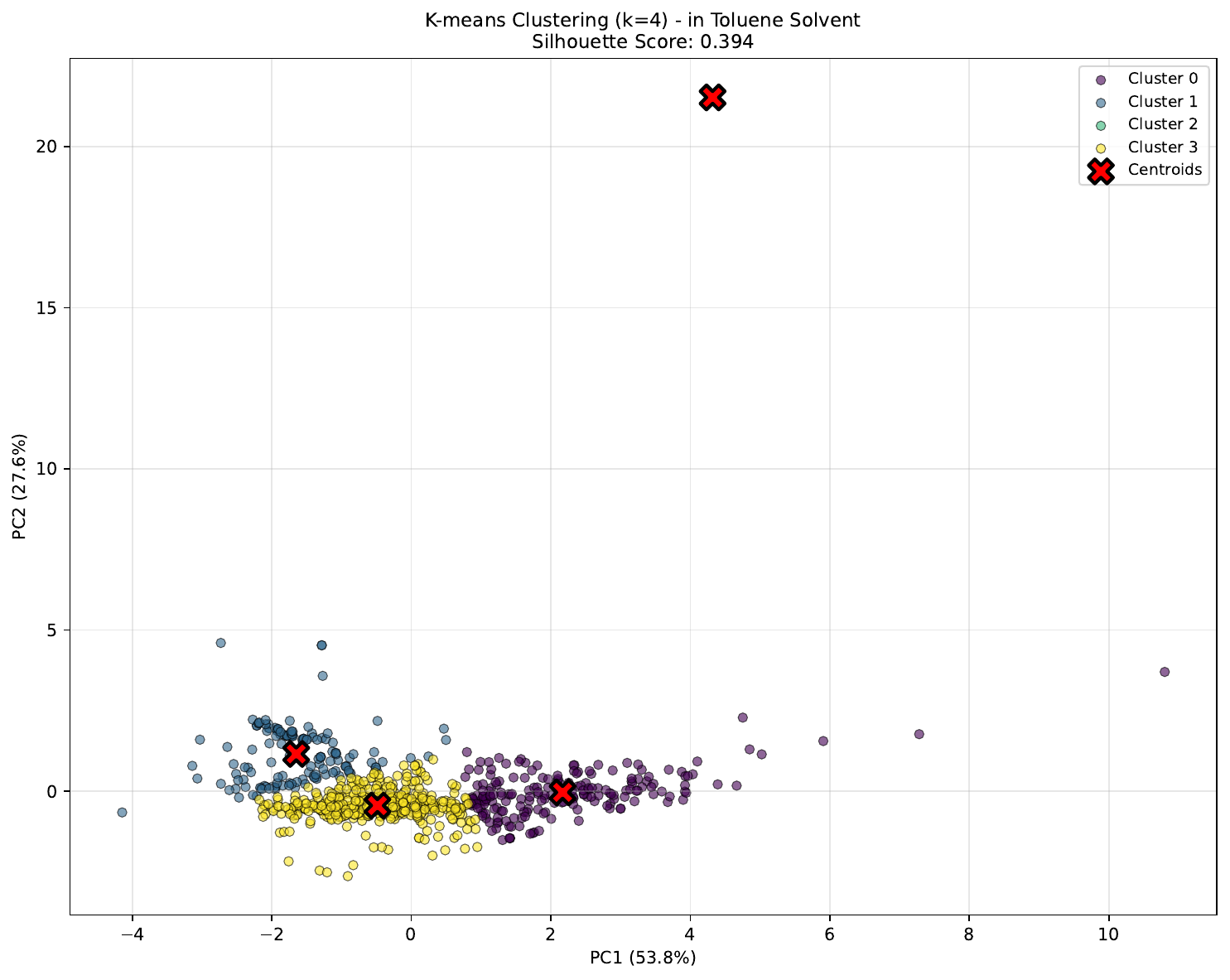}\\(b)
 \end{minipage}
\caption{K-means clustering (k=4) of the \num{747} TADF emitters based on their standardized photophysical properties in both the gas phase (a) and toluene 
solvent (b). The analysis partitions the data set into distinct families, most notably identifying a cluster of high-efficiency candidates (Cluster 1) with 
desirable properties for TADF applications.}
\label{fig:clustering}
\end{figure}

To complement the structure-based classification, we performed an unsupervised K-means clustering based on the molecules' photophysical properties. This 
data-driven approach identifies four distinct molecular families, as shown in \Cref{fig:clustering}:
\begin{itemize}
    \item \textbf{Cluster 0 (60\%, n=449): "Standard TADF"}. These molecules exhibit
moderate \deltaest~ (mean \qty{0.29}{\electronvolt}) and visible emission
(\qtyrange{500}{600}{\nano\meter}), representing the bulk of typical D-A and D-A-D
systems.

    \item \textbf{Cluster 1 (17\%, n=125): "High-Efficiency Candidates"}. This group is
characterized by a small \deltaest~ (mean \qty{0.18}{\electronvolt}), strong emission ($f
= 0.62$), and blue-green wavelengths (\qtyrange{450}{520}{\nano\meter}). It is
significantly enriched in optimized D-A-D and MR-TADF systems.

    \item \textbf{Cluster 2 (23\%, n=172): "NIR/Deep-Red Emitters"}. These molecules have
larger \deltaest~ (mean \qty{0.42}{\electronvolt}) and long-wavelength emission ($
\qty{>600}{\nano\meter}$), typically corresponding to architectures with extended
conjugation.

    \item \textbf{Cluster 3 ($<$1\%, n=1): Outliers}. A single outlier with extreme properties requiring
individual investigation.
\end{itemize}
This analysis confirms that while structural classification is useful, the photophysical property space naturally segregates molecules into performance-based 
families, with a clear cluster of high-potential candidates emerging directly from the data.

\subsection{Multi-resonance Tadf: a Distinct Paradigm for Blue Emitters}

Our analysis identifies Multiresonance (MR) emitters as a separate class, concentrated in the high-efficiency cluster. Unlike conventional D-A systems
that rely on through-space charge separation via torsional strain, MR-TADF emitters
achieve a small \deltaest~ through resonance effects in planar, rigid frameworks
\cite{Hatakeyama2016}. Through this mechanism, MR emitters can combine a small gap with high oscillator strength ($f$), enabling narrow-band, deep-blue emission 
($\text{FWHM} <
\qty{30}{\nano\meter}$) that is challenging for traditional D-A designs.
\Cref{tab:mr_vs_da} provides a quantitative comparison, showing that MR emitters achieve
small \deltaest~ values comparable to D-A-D systems but with significantly higher
oscillator strength. Complete molecular structures (SMILES) and classifications are
provided in the Supporting Information (Table S1).

However, as noted in the methodology, it is important to interpret these results with caution. The inherent electronic complexity of MR-TADF systems makes them 
prone to multi-reference character \cite{Chen2017}, which is not fully captured by the single-reference nature of the semiempirical methods used in this HTS. 
These results guide compound ranking but require higher-level validation for quantitative accuracy.

\begin{table}[!htbp]
    \caption{Photophysical property comparison of MR-TADF with conventional D-A systems,
calculated with sTDA-xTB in toluene. MR emitters achieve a small \deltaest~ comparable to
D-A-D systems but with significantly higher oscillator strength and blue-shifted emission,
highlighting a different design strategy.}
    \label{tab:mr_vs_da}
    \begin{tabularx}{\linewidth}{m{4cm}@{}S[table-format=3.2(2)] S[table-format=3.2(2)]
S[table-format=3.2(2)]}
        \toprule
        \textbf{Property}                                 & {\textbf{MR-TADF (n=67)}} &
{\textbf{D-A-D (n=182)}} & {\textbf{Simple D-A (n=248)}} \\ \midrule
        \deltaest~ [\unit{\electronvolt}]                    & 0.25 \pm 0.11             &
0.28 \pm 0.16            & 0.36 \pm 0.22                 \\
        Oscillator Strength ($f$)                         & 0.61 \pm 0.58             &
0.40 \pm 0.52            & 0.51 \pm 0.58                 \\
        Emission $\lambda_{\text{PL}}$ [\unit{\nano\meter}] & 468 \pm 52                &
512 \pm 68               & 528 \pm 84                    \\ \bottomrule
    \end{tabularx}
    \vspace{1em}
\end{table}

\subsection{Deconstructing Performance: Geometric and Electronic Design Levers}

\subsubsection{Torsional Angle as a Primary Control Parameter}

\begin{figure}[!ht]
\centering
\includegraphics[width=0.8\textwidth]{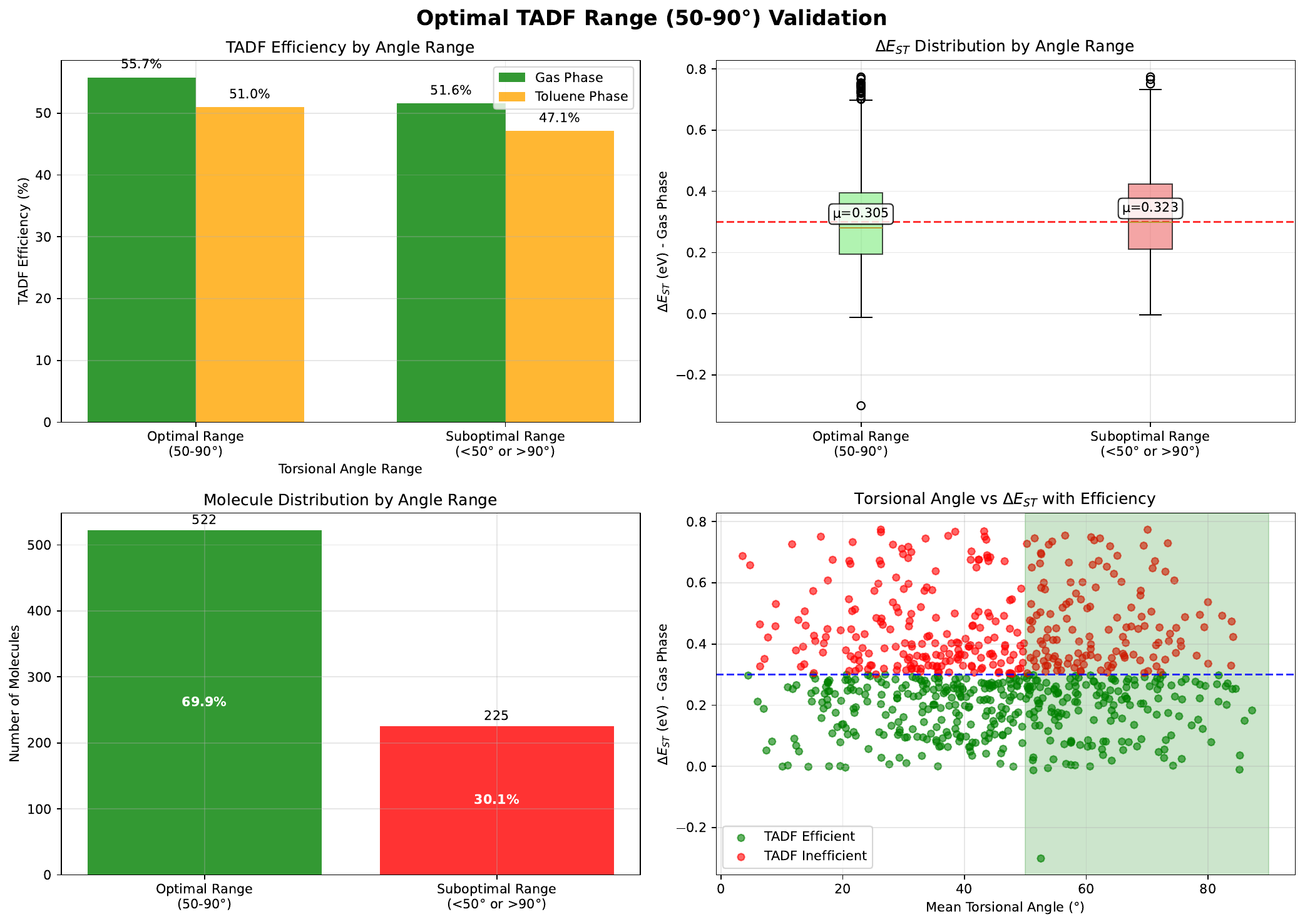}
\caption{Validation of the optimal torsional angle design rule. The bar chart shows the
percentage of molecules that are efficient TADF emitters (\deltaest~ $<$
\qty{0.3}{\electronvolt}) within the optimal (\qtyrange{50}{90}{\degree}) versus
suboptimal torsional angle ranges. Molecules in the optimal range exhibit a statistically
significant ($p < 0.001$) increase in TADF efficiency.}
\label{fig:tadf_design_rules}
\end{figure}

For conventional D-A architectures, our large-scale analysis provides statistical
validation for the central role of the D-A torsional angle. We confirm that an optimal
window exists in the \qtyrange{50}{90}{\degree} range (\Cref{fig:tadf_design_rules}).
This twisted geometry is the primary lever for minimizing the exchange integral by
physically separating the HOMO (donor-localized) and LUMO (acceptor-localized), thereby
enforcing the charge-transfer character of the $S_1$ state, a finding consistent with
previous targeted studies \cite{Olivier2017}.

\subsubsection{Quantifying the Torsion-induced Electronic Trade-off}

\begin{figure}[!ht]
\centering
\includegraphics[width=0.9\textwidth]{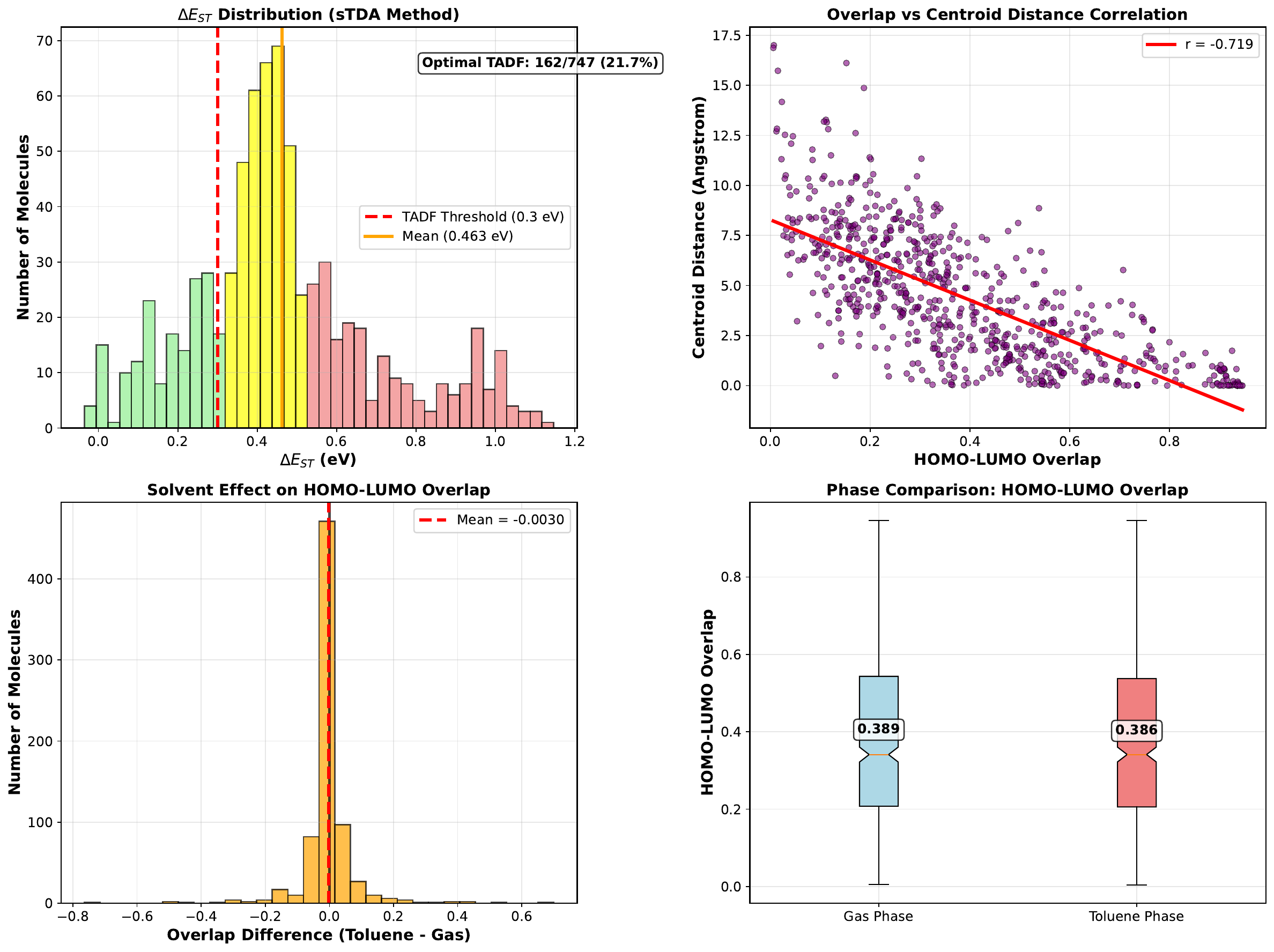}
\caption{Correlation analysis between HOMO-LUMO overlap and key photophysical properties.
The strong negative correlation with \deltaest~ (left) validates the spatial separation
model for TADF design, while the strong positive correlation with oscillator strength
(right) highlights the fundamental design trade-off that molecular engineers must
navigate.}
\label{fig:correlation_analysis}
\end{figure}

This geometric control is directly reflected in the electronic structure. As shown in
\Cref{fig:correlation_analysis}, twisting the molecule reduces the HOMO-LUMO overlap
($S'_{\text{HL}}$), which correlates strongly with a smaller \deltaest~ (Spearman $\rho =
-0.45$). However, this comes at a cost: the oscillator strength, essential for a high
radiative rate, is directly proportional to this overlap ($\rho = 0.78$). This
visualization makes the fundamental TADF design trade-off explicit: minimizing the energy
gap via orbital separation inherently suppresses emission intensity. An optimal region for
high-performance emitters lies where there is intermediate overlap (\numrange{0.15}{0.30})
that balances these competing requirements.

To move beyond this qualitative picture and establish a more rigorous quantitative link,
future work could employ methods such as Natural Transition Orbitals (NTOs) or the RespA
procedure \cite{grimme2019exploration, Wergifosse2024a}. These tools would allow for a
compact representation of the virtual excitations and rigorously connect the molecular
geometry to the precise charge-transfer character (e.g., via centroid distance $\Delta r$)
and the ensuing coupling efficiency \cite{Etienne2014a}.

\subsection{Implications for Reverse Intersystem Crossing ($k_{\text{risc}}$)}

While a small \deltaest~ is the primary thermodynamic requirement for TADF, the overall
efficiency is governed by the kinetics of RISC. While direct calculation of the
$k_{\text{RISC}}$ rate is outside the scope of HTS, the derived structural principles must
be explicitly connected to the kinetics of TADF. Within the Marcus theory framework,
$k_{\text{RISC}}$ is proportional to the square of the spin-orbit coupling (SOC) matrix
element and exponentially dependent on \deltaest~ and the reorganization energy
($\lambda$) \cite{Marcus-RISC-Theory-1985}.

According to El-Sayed's rules, SOC is maximized when the interconverting singlet ($S_1$)
and triplet ($T_1$) states have different orbital characters (e.g., charge-transfer vs.
local-excitation, $^1$CT $\leftrightarrow$ $^3$LE) \cite{El-Sayed-Rules-1968}. This
creates a mechanistic challenge: the very act of creating a pure CT state in $S_1$ to
minimize \deltaest~ can lead to a $T_1$ state that also has pure CT character, making the
transition spin-forbidden ($^1$CT $\leftrightarrow$ $^3$CT) and suppressing
$k_{\text{RISC}}$.

Molecular design can resolve this tension. A torsional angle of \qtyrange{50}{90}{\degree} strikes the necessary balance. It is twisted enough to enforce 
significant CT character in $S_1$ and a small
\deltaest, but not so orthogonal as to completely eliminate mixing with local-excited (LE)
states. This promotes a necessary divergence in electronic character between the $S_1$
(primarily CT) and $T_1$ (often LE or mixed CT/LE) states, ensuring a non-zero SOC and
satisfying both the thermodynamic and kinetic requirements for efficient RISC.
Furthermore, the incorporation of heavy atoms such as sulfur into donor units is a known
strategy to intrinsically enhance SOC, which can compensate for slightly larger \deltaest~
values. \Cref{tab:risc_descriptors} summarizes how key structural parameters influence the
factors governing $k_{\text{RISC}}$.

{
\renewcommand{\arraystretch}{1.25}
\begin{table}[!ht]
\caption{Connection between molecular descriptors and the physical factors governing the RISC rate ($k_{\text{RISC}}$).}
\label{tab:risc_descriptors}
\small
\begin{tabularx}{\linewidth}{>{\raggedright\arraybackslash}m{0.25\textwidth}
>{\raggedright\arraybackslash}m{0.36\textwidth}
>{\raggedright\arraybackslash}m{0.35\textwidth}}
\toprule
\textbf{Molecular descriptor}                 & \textbf{Influence on RISC factor}
                    & \textbf{Optimal range/strategy} \\ \midrule
HOMO-LUMO overlap                             & Directly modulates \textbf{\deltaest}
(thermodynamics)         & \numrange{0.15}{0.30}           \\
D-A torsional angle                           & Balances \textbf{\deltaest} and
\textbf{SOC} (kinetics)        & \qtyrange{50}{90}{\degree}      \\
Heteroatoms (S, Se)                           & Directly increase \textbf{SOC} via
heavy-atom effect           & Incorporate S-based donors      \\
\parbox[c]{0.25\textwidth}{Conformational rigidity (e.g., in MR systems)} & Minimizes
reorganization energy $\mathbf{\lambda}$ (kinetics) & Use bridged or fused structures \\
\bottomrule
\end{tabularx}
\end{table}
}

\subsection{Synthesis of a Data-driven Design Strategy}

The combination of these large-scale statistical analyses allows us to formulate a set of quantitative, data-driven design guidelines for accelerating the 
discovery of novel TADF emitters, summarized in \Cref{tab:design_guidelines}. These rules provide a clear roadmap for molecular engineering, from selecting a 
high-performance architecture to fine-tuning its properties via geometric and electronic modifications.

{
\renewcommand{\arraystretch}{1.25}
\begin{table}[!ht]
\caption{Quantitative TADF design guidelines derived from the high-throughput screening.}
\label{tab:design_guidelines}
\begin{tabularx}{\linewidth}{>{\raggedright\arraybackslash}m{0.23\textwidth}
>{\raggedright\arraybackslash}m{0.25\textwidth}m{0.45\textwidth}}
\toprule
\textbf{Parameter}                   & \textbf{Target range / strategy} &
\textbf{Rationale}                                                             \\ \midrule
Architecture                         & D-A-D or MR                      &
\parbox[c]{0.45\textwidth}{Statistically superior performance for general or blue
emission, respectively.} \\
D-A Torsional Angle                  & \qtyrange{50}{90}{\degree}       & Optimal balance
between small \deltaest~ and non-zero SOC.                    \\
HOMO-LUMO Overlap ($S'_{\text{HL}}$) & \numrange{0.15}{0.30}            & Navigates the
trade-off between \deltaest~ and oscillator strength.           \\
Oscillator Strength ($f_{S_1}$)      & $\num{> 0.1}$                          & Ensures a
high radiative decay rate for bright emission.                       \\ \bottomrule
\end{tabularx}
\end{table}
}

\subsection{Future Outlook and Data Exploitation}\label{sec:data_ml}

\subsubsection{The Hts Data Set as a Machine Learning Substrate}
The 747-molecule data set, computed with a validated protocol \cite{TchapetArxivA1-2025}, constitutes the main output of this study. This data set, encompassing 
key geometric (torsional angles), electronic ($S'_{\text{HL}}$, \deltaest, $f$), and structural descriptors, represents an ideal foundation for Machine Learning 
(ML) approaches \cite{Friederich2021ML, Lee2021ML}. The low-dimensional nature of the property space (\qty{88.8}{\percent} of variance captured by three 
components) confirms its suitability for training Quantitative Structure-Property Relationship (QSPR) models. These data can train ML models to predict 
properties like $k_{\text{RISC}}$ or $\Phi_{\text{PL}}$. Such models would reduce the need for expensive calculations.

\subsubsection{Actionable Design Targets}
By applying the derived quantitative design rules (e.g., optimal torsion range, D-A-D architecture) and the performance criteria (\deltaest $< 
\qty{0.1}{\electronvolt}$ and oscillator strength $f > 0.1$), our HTS successfully filtered the expansive chemical space to identify \num{127} high-priority 
candidate molecules. The full candidate list appears in the Supporting Information. These molecules are proposed as targets for synthesis and testing.

\subsection{Sensitivity Analysis of Design Guidelines}

To assess the robustness of our HTVS-derived design guidelines, we performed a comprehensive sensitivity analysis examining how variations in threshold choices 
affect the identification of promising TADF candidates. This analysis addresses the inherent subjectivity in defining quantitative cutoffs and provides 
confidence intervals for our screening criteria.

\subsubsection{Threshold Sensitivity for \deltaest~ Criteria}

We systematically varied the \deltaest~ threshold from 0.2 to 0.4 eV to examine its impact on candidate selection. The results demonstrate that threshold 
choice 
significantly influences the number of molecules classified as promising: \deltaest~ $\leq 0.2$ eV identifies 163 molecules (21.8\%), while \deltaest $\leq 
0.4$ eV captures 507 molecules (67.9\%). The 0.3 eV threshold employed throughout this study represents a balanced choice that captures 345 molecules (46.2\%) 
while maintaining selectivity.

When combining \deltaest~ and HOMO-LUMO overlap criteria, the sensitivity becomes more pronounced. For example, applying both \deltaest $\leq 0.3$ eV and 
overlap $\leq 0.4$ yields 272 molecules (36.4\%), while the more restrictive combination identifies 134 molecules (17.9\%). This demonstrates that our 
guidelines represent screening criteria rather than absolute predictors.

Importantly, the relative ranking of molecular architectures remains consistent across different threshold choices. D-A-D architectures consistently outperform 
other motifs regardless of the specific \deltaest~ cutoff employed, indicating that our architectural guidelines are robust to threshold variations. Similarly, 
the favorable torsional angle range of \qtyrange{50}{90}{\degree} maintains its statistical significance across different analysis parameters.

This sensitivity analysis reveals that while the absolute number of promising candidates varies with threshold choice, the underlying structure-property 
relationships remain consistent. Our guidelines should be viewed as HTVS-derived screening criteria that provide relative rankings and identify promising 
synthetic targets, rather than quantitative predictors of absolute TADF performance. The robustness of trends across different thresholds supports the 
mechanistic validity of our structure-property relationships while acknowledging the inherent limitations of computational screening approaches.

\subsection{Validation Against Experimental Data}\label{sec:validation}

To assess the predictive accuracy of our computational protocol, we compiled an extensive experimental validation dataset from the literature. We systematically 
extracted singlet-triplet energy gap ($\Delta E_{\mathrm{ST}}$) measurements from a curated database of TADF molecules,\cite{Huang_2024} focusing on compounds 
with well-documented experimental characterization. After matching molecular structures via SMILES identifiers and filtering for data quality, we obtained 
\num{379} molecules with both experimental $\Delta E_{\mathrm{ST}}$ values and corresponding calculated descriptors from our workflow.

The experimental dataset spans a broad range of molecular architectures, including Multi-D/A ($n$=136), 2D-0A ($n$=68), 1D-0A ($n$=53), D-A-D ($n$=28), 2D-2A 
($n$=26), A-D-A ($n$=15), and D-A ($n$=12) systems, among others. This diversity enables a comprehensive assessment of our method's performance across different 
design strategies. All experimental values were extracted from peer-reviewed publications with traceable DOI references (see Supporting Information Table S2 for 
the complete dataset).

\begin{figure}[htbp]
    \centering
    \includegraphics[width=0.95\textwidth]{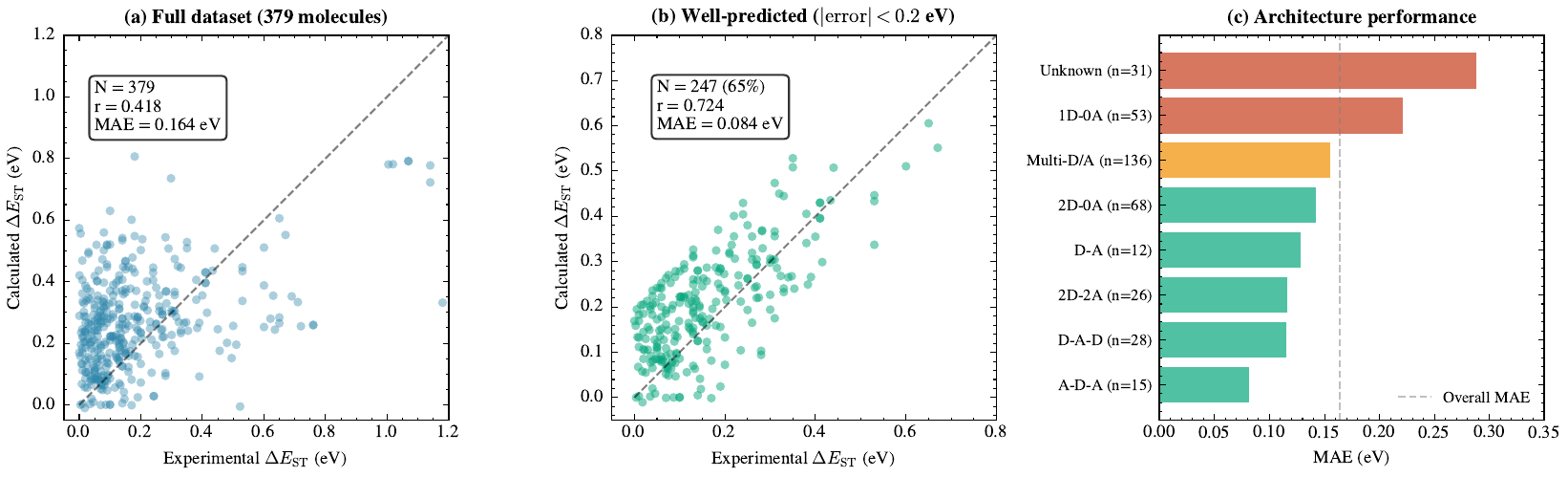}
    \caption{Validation of calculated $\Delta E_{\mathrm{ST}}$ values against experimental data for the \num{379}-molecule benchmark set. (a) Full dataset 
showing moderate correlation ($r$ = 0.418, MAE = \qty{0.164}{\electronvolt}). (b) Well-predicted subset with |error| $<$ \qty{0.2}{\electronvolt} ($n$ = 247, 
65\%) showing strong correlation ($r$ = 0.724, MAE = \qty{0.084}{\electronvolt}). (c) Architecture-dependent performance, with A-D-A, D-A-D, and 2D-2A showing 
the lowest prediction errors. The dashed lines in (a) and (b) indicate perfect agreement. Complete dataset with DOI references provided in Supporting 
Information Table S2.}
    \label{fig:validation}
\end{figure}

\Cref{fig:validation} presents the correlation between experimental and calculated $\Delta E_{\mathrm{ST}}$ values for the full \num{379}-molecule dataset. The 
gas-phase sTDA calculations yield a Pearson correlation coefficient of $r$ = \num{0.418} ($p$ = \num{1.8e-17}), indicating a statistically significant positive 
correlation. The mean absolute error (MAE) is \qty{0.164}{\electronvolt}, with a systematic overestimation bias of \qty{+0.086}{\electronvolt}. While this 
correlation is moderate, it demonstrates that our semi-empirical approach captures the general trends in singlet-triplet splitting across diverse TADF 
architectures.

The scatter in the correlation reflects several factors. First, experimental $\Delta E_{\mathrm{ST}}$ values are typically derived from photophysical 
measurements in solution or thin films, whereas our calculations employ gas-phase or implicit solvent models. Environmental effects---including host-guest 
interactions, conformational distributions, and specific solvation---can significantly influence the measured singlet-triplet gap. Second, the sTDA method, 
while computationally efficient, approximates the excited-state electronic structure and may struggle with certain molecular motifs, particularly those 
involving charge-transfer states with near-degeneracies or strong vibronic coupling.

To identify the subset of molecules for which our protocol performs reliably, we analyzed the error distribution. Approximately 65\% of the dataset (\num{247} 
molecules) exhibits prediction errors below \qty{0.2}{\electronvolt}. For this well-predicted subset, the correlation improves substantially to $r$ = 
\num{0.724} ($p$ = \num{2.2e-41}), with a reduced MAE of \qty{0.084}{\electronvolt} (\Cref{fig:validation}b). This suggests that for the majority of TADF 
emitters, our workflow provides quantitatively useful predictions. The remaining 35\% of molecules with larger errors warrant further investigation; preliminary 
analysis indicates that many of these outliers involve architectures with multiple conformational minima or acceptor units prone to sTDA artifacts.

We further examined the performance of our method across different molecular architectures (\Cref{fig:validation}c). The A-D-A architecture shows the lowest MAE 
(\qty{0.082}{\electronvolt}, $n$=15), followed by D-A-D (\qty{0.115}{\electronvolt}, $n$=28) and 2D-2A (\qty{0.116}{\electronvolt}, $n$=26). Multi-D/A systems, 
which constitute the largest subset, exhibit a moderate MAE of \qty{0.156}{\electronvolt}. In contrast, 1D-0A architectures present a greater challenge, with an 
MAE of \qty{0.221}{\electronvolt} ($n$=53). This architecture-dependent performance likely reflects differences in the electronic structure complexity and the 
extent of charge-transfer character in the excited states.

Outliers with prediction errors exceeding \qty{0.4}{\electronvolt} account for approximately 6\% of the dataset (\num{23} molecules). Manual inspection reveals 
that several of these cases involve molecules with sulfonyl acceptors or extended conjugated systems where conformational flexibility may lead to discrepancies 
between the calculated ground-state geometry and the emissive conformer in the experimental environment. Others correspond to molecules with very small 
experimental $\Delta E_{\mathrm{ST}}$ values ($<$\qty{0.05}{\electronvolt}), where even minor computational inaccuracies translate to large relative errors.

Despite these limitations, the \num{379}-molecule validation demonstrates that our computational protocol achieves reasonable predictive accuracy for a 
substantial fraction of the TADF design space. The statistically significant correlation, combined with the strong performance for the well-predicted subset, 
supports the use of our descriptors for virtual screening and structure-property relationship analysis. For practical applications, we recommend prioritizing 
candidates from architectures with demonstrated low prediction errors (A-D-A, D-A-D, 2D-2A) and exercising caution with 1D-0A systems or molecules featuring 
structural motifs associated with larger deviations.

\subsection{Spin-orbit Coupling Validation}\label{sec:soc_validation}

To validate the structural proxies employed in our kinetic analysis (Section~\ref{sec:methods}), we calculated direct S$_1$--T$_1$ spin-orbit coupling (SOC) 
matrix elements for 17 representative molecules using TD-DFT with full SOC treatment ($\omega$B97X-D4/def2-TZVP, ORCA 6.1.0\cite{Neese2022}). The molecules span 
diverse architectural classes (D-A, D-A-D, MR-TADF, spiro systems) and cover a wide range of HOMO-LUMO overlaps (0.061--0.740) and \deltaest~ values 
(0.051--0.486~eV).

The calculated T$_1$--S$_1$ SOC values for organic TADF emitters ranged from 0.058~cm$^{-1}$ (BACN) to 14.025~cm$^{-1}$ (DMAC-BP), with a median of 
0.490~cm$^{-1}$ (excluding the platinum complex PtOEP). These values are consistent with literature reports for organic TADF systems, which typically exhibit 
SOC in the range of 0.1--10~cm$^{-1}$.\cite{Gibson2016,Penfold2018} The platinum-containing molecule PtOEP showed significantly higher SOC (3.021~cm$^{-1}$), 
reflecting the well-known heavy atom effect that enhances spin-orbit coupling.\cite{Yersin2011}

Correlation analysis between SOC matrix elements and structural descriptors revealed weak linear relationships: HOMO-LUMO overlap ($r$ = 0.037, $p$ = 0.89), 
\deltaest~ ($r$ = 0.067, $p$ = 0.80), and oscillator strength ($r$ = -0.182, $p$ = 0.48). The absence of strong correlations is physically meaningful and 
reflects the complex dependence of SOC on orbital angular momentum, nodal structure, and local symmetry---factors that cannot be captured by simple geometric or 
energetic descriptors alone.\cite{ElSayed1968,Marian2012}

These findings validate our approach of using structural descriptors (torsional angles, HOMO-LUMO overlap) as \textit{qualitative proxies} for favorable SOC in 
high-throughput screening, while acknowledging that quantitative SOC predictions require explicit quantum chemical calculations. The weak correlations also 
explain why direct SOC calculations are essential for detailed kinetic modeling, even though structural guidelines can effectively identify promising TADF 
candidates for experimental synthesis.

Complete SOC data for all 17 molecules, including X, Y, and Z components of the coupling matrix elements, are provided in Supporting Information Table~S7 and 
Figure~S8.

\subsection{Multi-reference Tadf Validation}\label{sec:mr_tadf_validation}

For MR-TADF emitters specifically, the single-reference nature of sTD-DFT-xTB represents a fundamental limitation for quantitative energy predictions. As 
demonstrated in our TD-DFT validation (Supporting Information Table S6), absolute \deltaest~ values for MR-TADF molecules (24PQ-Cz, phenazasiline) show large 
deviations from higher-level theory (up to \qty{1.4}{\electronvolt} for individual molecules). This is expected given the multi-reference character of these 
systems arising from extensive $\pi$-conjugation and multiple resonance stabilization.\cite{Hatakeyama2016,Kondo2019}

However, the relative ordering and qualitative trends are preserved between sTD-DFT-xTB and TD-DFT (Table S6), with both methods ranking phenazasiline as having 
smaller \deltaest~ than 24PQ-Cz. This trend preservation is consistent with literature precedent,\cite{Olivier2022,Gomez-Bombarelli2016} which shows that 
approximate single-reference methods remain valid for high-throughput screening and candidate prioritization, even for systems with multi-reference character. 
The key insight is that screening methods need not provide quantitative accuracy to be useful; they must reliably identify promising candidates for further 
investigation.

For the MR-TADF molecules identified through our screening protocol, we recommend systematic post-hoc validation before experimental synthesis. Specifically:
\begin{itemize}
    \item Single-reference correlation methods (SCS-CC2, ADC(2)) for initial refinement of \deltaest~ predictions
    \item Multi-reference methods (STEOM-DLPNO-CCSD, CASPT2, NEVPT2) for molecules with predicted \deltaest~ $< \qty{0.2}{\electronvolt}$
    \item Assessment of multi-reference diagnostics (T$_1$, D$_1$) to identify systems requiring specialized treatment
\end{itemize}

This hierarchical validation approach balances computational cost with accuracy, leveraging the efficiency of sTD-DFT-xTB for initial screening while 
acknowledging the need for higher-level methods for quantitative predictions of specific candidates. Future development of multi-reference-aware descriptors or 
machine learning models trained on post-HF data could further improve screening accuracy for MR-TADF systems.

\section{CONCLUSIONS}\label{sec:conclusions}

In this work, we performed a large-scale computational screening of \num{747} TADF emitters to extract quantitative structure-property relationships for guiding 
molecular design. By analyzing architecture, geometry, and electronic structure across 747 molecules, we derived design guidelines grounded in both 
thermodynamic and kinetic considerations.

We have quantitatively
demonstrated the superiority of the D-A-D architectural motif and identified an optimal
donor-acceptor torsional angle window of \qtyrange{50}{90}{\degree}. This optimal geometry does more than minimize \deltaest; it also preserves sufficient 
spin-orbit coupling for RISC, satisfying both thermodynamic and kinetic requirements. Furthermore, our clustering results establish MR-TADF as a distinct class, 
capable of combining the small \deltaest{} and high $f$ needed for blue emission.

The design guidelines, candidate list, and mechanistic analysis reported here should facilitate further TADF development. Experimental characterization of the 
identified candidates would test these predictions.

\section*{Author Contributions}
J.-P.T.N.: Conceptualization, Methodology, Software, Formal Analysis, 
Investigation, Data Curation, Writing – Original Draft, Visualization.\\ 
E.V.K.T.: Writing – Review \& Editing. \\
A.M.:  Writing – Review \& Editing. \\
S.G.N.E.: Supervision, Writing – Review \& Editing.

\section*{Notes}
An AI assistant (Claude, Anthropic) was used during manuscript preparation to assist with language editing, formatting, and structural organization. All 
scientific content, data analysis, interpretations, and conclusions are entirely the work of the authors. The authors have critically reviewed and verified all 
text, and take full responsibility for the accuracy of the manuscript.

\section*{Acknowledgements}

We gratefully acknowledge \textbf{Dr. Benjamin Panebei Samafou} for his generous support
and provision of computational resources, which were instrumental in enabling the
numerical simulations and data analyses presented in this work.

\section*{Data and Software Availability}
The computational data set of 747 TADF emitters, including optimized 
geometries, excited-state properties, and analysis scripts, is freely 
available at [Zenodo DOI: 10.5281/zenodo.17436069] / [GitHub:
\url{https://github.com/TchapetNjafa/Result_article1_TADF_xTB/tree/main}].
All calculations used the xtb package (version 6.7.1) available at 
\url{https://github.com/grimme-lab/xtb}, the CREST package (version 3.0.2) available at
\url{https://github.com/crest-lab/crest},
the xtb4stda package (version 1.0) available at
\url{https://github.com/grimme-lab/xtb4stda}, the stda package (version 1.6.3) available
at \url{https://github.com/grimme-lab/std2}
and Multiwfn package (version 3.8(dev)) available at \url{http://sobereva.com/multiwfn}.

\section*{Conflicts of Interest}
The authors declare no competing financial interests.

\section{Associated content}

\subsection{Supporting information}

The Supporting Information is available free of charge at
\url{https://pubs.acs.org/doi/...}

\noindent This Supporting Information contains:
\begin{itemize}
\item \Cref{tab:architecture_analysis}: Architectural classification and performance statistics for TADF emitters
\item \Cref{tab:validation}: Experimental validation data for representative molecules
\item \Cref{tab:mr_tadf_validation}: Multi-reference TADF validation - xTB vs TD-DFT comparison
\item \Cref{tab:soc_summary}: Spin-orbit coupling matrix elements for 17 representative molecules
\item \Cref{fig:adiabatic_validation}: Correlation analysis of SOC with structural descriptors
\item Additional computational details and analysis parameters
\end{itemize}

{\renewcommand{\arraystretch}{2.25}
\begin{longtable}[H]{l w{c}{6.5cm} *{3}{c}S[table-format=4.3]S[table-format=2.3]}
\caption{Molecular architecture and classification of the studied TADF emitters. This table provides the SMILES representation of each molecule, along with its structural classification.} 
\label{tab:architecture_analysis} \\
\toprule 
Molecule Id & SMILES & Total donors & Total acceptors & Architecture & {Molecular weight}
& {logp} \\
\midrule
\endfirsthead 
\multicolumn{7}{c}{{\tablename} \thetable{} -- Continued from previous page} \\ 
\toprule 
Molecule Id & SMILES & Total donors & Total acceptors & Architecture & {Molecular weight}
& {logp} \\
\midrule 
\endhead 
\midrule 
\multicolumn{7}{r}{{Continued on next page}} \\ 
\endfoot 
\bottomrule 
\endlastfoot 
4CzTPN & \smiles{N\#Cc1c(-n2c3ccccc3c3ccccc32)c(-n2c3ccccc3c3ccccc32)c(C\#N)c(-n2c3ccccc3c3ccccc32)c1-n1c2ccccc2c2ccccc21} & 4 & 4 & Multi-D/A & 788.914 & 13.818 \\
DMAC-DPS & \smiles{CC1(C)c2ccccc2N(c2ccc(S(=O)(=O)c3ccc(N4c5ccccc5C(C)(C)c5ccccc54)cc3)cc2)c2ccccc21} & 2 & 0 & 2D-0A & 632.829 & 10.738 \\
PSPCz & \smiles{C1=CC=C(C=C1)S(=O)(=O)C2=CC=CC(=C2)N3C4=CC=CC=C4C5=CC=CC=C53} & 1 & 0 & 1D-0A & 383.472 & 5.617 \\
Px2BP & \smiles{C1=CC=C2C(=C1)N(C3=CC=CC=C3O2)C4=CC=C(C=C4)C(=O)C5=CC=C(C=C5)N6C7=CC=CC=C7OC8=CC=CC=C86} & 4 & 0 & Multi-D/A & 544.610 & 10.069 \\
CzS2 & \smiles{C1=CC=C2C(=C1)C3=CC=CC=C3N2C4=CC=C(C=C4)S(=O)(=O)C5=CC=C(C=C5)N6C7=CC=CC=C7C8=CC=CC=C86} & 2 & 0 & 2D-0A & 548.667 & 8.714 \\
2TCz-DPS & \smiles{CC(C)(C)C1=CC2=C(C=C1)N(C3=C2C=C(C=C3)C(C)(C)C)C4=CC=C(C=C4)S(=O)(=O)C5=CC=C(C=C5)N6C7=C(C=C(C=C7)C(C)(C)C)C8=C6C=CC(=C8)C(C)(C)C} & 2 & 0 & 2D-0A & 773.099 & 13.904 \\
TDBA-DI & \smiles{CC(C)(C)c1ccc2c(c1)B1c3cc(C(C)(C)C)ccc3Oc3cc(-n4c5ccccc5c5c6c(c7ccccc7n6-c6ccccc6)c6c(c7ccccc7n6-c6ccccc6)c54)cc(c31)O2} & 3 & 0 & Multi-D/A & 877.897 & 14.300 \\
OPDPO & \smiles{O=CC1=CCC(c2ccccc2N2c3ccccc3Sc3ccccc32)(P(=O)(c2ccccc2)c2ccccc2)C=C1} & 2 & 0 & 2D-0A & 581.677 & 8.916 \\
\end{longtable}
}

\section{Experimental Validation Data}

\scriptsize
\setlength{\tabcolsep}{2.5pt}
\begin{longtable}{llcccccc}
\caption{Experimental validation benchmark set (\num{379} molecules). Complete dataset comparing calculated (gas-phase screening) and experimental $\Delta E_{\mathrm{ST}}$ values across diverse TADF architectures.}
\label{tab:validation} \\
\toprule
\textbf{Molecule} & \textbf{Arch.} & \textbf{Exp.} & \textbf{Calc.} & \textbf{Error} & \boldmath{$f$} & \textbf{Overlap} & \textbf{Ref.} \\
 &  & \boldmath{$\Delta E_{\mathrm{ST}}$} & \boldmath{$\Delta E_{\mathrm{ST}}$} & \textbf{(eV)} &  &  &  \\
 &  & \textbf{(eV)} & \textbf{(eV)} &  &  &  &  \\
\midrule
\endfirsthead
\multicolumn{8}{c}{\tablename\ \thetable\ -- \textit{Continued from previous page}} \\
\toprule
\textbf{Molecule} & \textbf{Arch.} & \textbf{Exp.} & \textbf{Calc.} & \textbf{Error} & \boldmath{$f$} & \textbf{Overlap} & \textbf{Ref.} \\
 &  & \boldmath{$\Delta E_{\mathrm{ST}}$} & \boldmath{$\Delta E_{\mathrm{ST}}$} & \textbf{(eV)} &  &  &  \\
 &  & \textbf{(eV)} & \textbf{(eV)} &  &  &  &  \\
\midrule
\endhead
\midrule
\multicolumn{8}{r}{\textit{Continued on next page}} \\
\endfoot
\bottomrule
\endlastfoot
Diphenylsulfane & 0D-1A & 0.280 & 0.094 & -0.186 & 0.001 & 0.479 & \cite{doi101039D2TC01435A} \\
CzmPPC & 0D-1A & 0.280 & 0.103 & -0.177 & 0.001 & 0.472 & \cite{doi101039D2TC01435A} \\
TCbTrz & 0D-1A & 0.270 & 0.220 & -0.050 & 0.381 & 0.133 & \cite{doi101039C8NJ00767E} \\
APPT & 0D-1A & 0.070 & 0.186 & +0.116 & 0.011 & 0.340 & \cite{doi101016jdyepig2021109474} \\
PX-TRZ & 0D-1A & 0.056 & 0.242 & +0.186 & 0.031 & 0.256 & \cite{doi101039C9TC06497D} \\
DMIC-TRZ & 0D-1A & 0.056 & 0.519 & +0.463 & 0.231 & 0.778 & \cite{doi101016jorgel2022106536} \\
D--A--D & 0D-2A & 0.242 & 0.028 & -0.214 & 0.007 & 0.544 & \cite{doi101039D1TC05730H} \\
phenazasiline & 0D-2A & 0.242 & 0.028 & -0.214 & 0.007 & 0.545 & \cite{doi101039D1TC05730H} \\
\end{longtable}

\begin{figure}[!ht]
    \centering
    \includegraphics[width=0.85\textwidth]{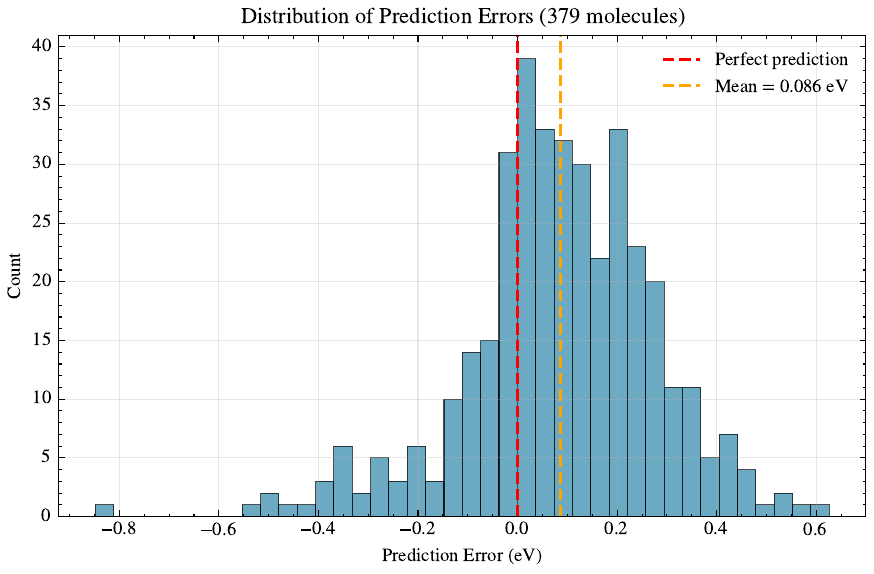}
    \caption{Distribution of prediction errors for the 379-molecule validation dataset. The histogram shows that most molecules have errors within $\pm$0.2 eV, 
with a mean error of +0.086 eV (systematic overestimation). The distribution is approximately normal with some outliers at larger absolute errors.}
    \label{fig:error_distribution}
\end{figure}

\begin{figure}[!ht]
    \centering
    \includegraphics[width=0.90\textwidth]{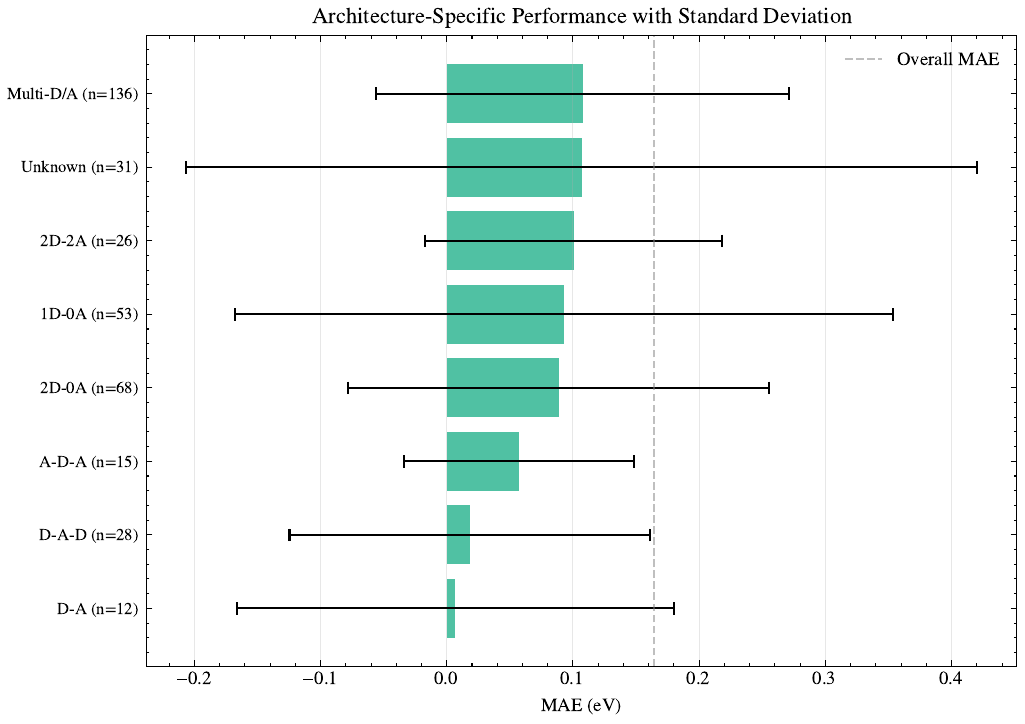}
    \caption{Architecture-specific performance with standard deviations. Error bars represent the variability in prediction accuracy within each architecture 
class. A-D-A and D-A-D architectures show the lowest mean absolute errors and smallest variability, while 1D-0A systems exhibit larger errors and greater 
variability, reflecting the challenges in predicting charge-transfer states for these motifs.}
    \label{fig:architecture_performance}
\end{figure}

\subsection{Outlier Analysis}

\begin{figure}[!ht]
    \centering
    \includegraphics[width=\textwidth]{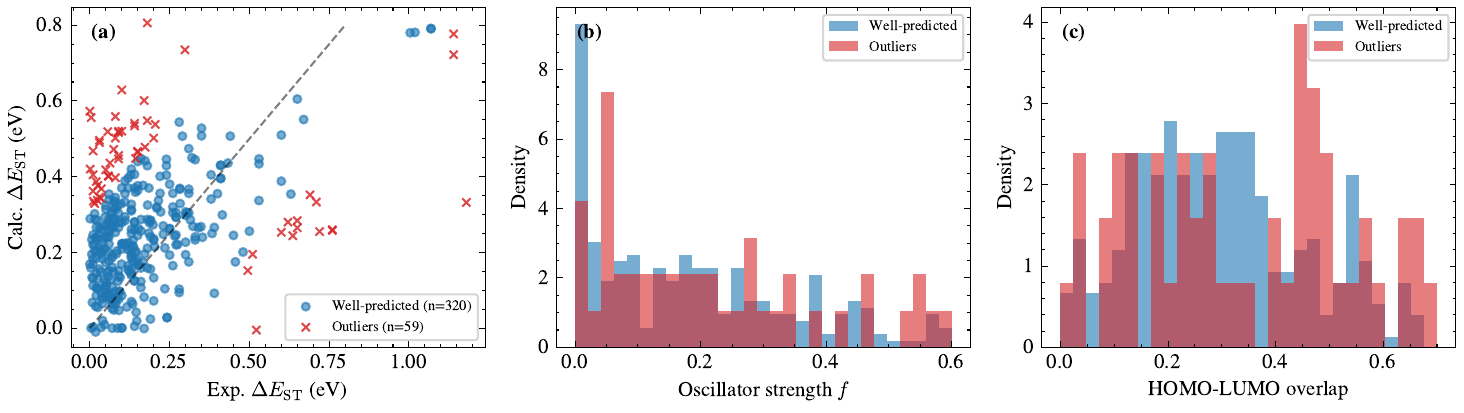}
    \caption{Outlier analysis of the 379-molecule validation set. (a) Scatter plot highlighting outliers (red crosses, $|$error$|$ > 0.3~eV) versus 
well-predicted molecules (blue circles). (b) Distribution of oscillator strength $f$ for outliers and well-predicted molecules. (c) Distribution of HOMO-LUMO 
overlap for both groups. Outliers show systematically higher overlap values, indicating stronger charge-transfer character.}
    \label{fig:SI_outlier_analysis}
\end{figure}

To understand the limitations of our computational approach, we analyzed the 59 molecules (15.6\% of the dataset) with prediction errors exceeding 0.3~eV. Figure~\ref{fig:SI_outlier_analysis} shows the distribution of these outliers and their characteristic descriptors.

\subsubsection{Error Patterns}

The outliers exhibit two distinct error patterns:
\begin{itemize}
    \item \textbf{Overestimation} (Calc $>$ Exp): 43 molecules (72.9\% of outliers), predominantly with very small experimental $\Delta E_{\mathrm{ST}}$ values 
($<$0.1~eV). The mean experimental value for this group is 0.04~eV, while calculations predict 0.41~eV, suggesting systematic overestimation for near-zero 
singlet-triplet gaps.
    
    \item \textbf{Underestimation} (Calc $<$ Exp): 16 molecules (27.1\% of outliers), primarily with large experimental $\Delta E_{\mathrm{ST}}$ values 
($>$0.5~eV). The mean experimental value is 0.72~eV versus calculated 0.32~eV, indicating difficulty in predicting molecules with large singlet-triplet 
splittings.
\end{itemize}

\subsubsection{Descriptor Analysis}

Outliers show distinct characteristics compared to well-predicted molecules:

\begin{itemize}
    \item \textbf{HOMO-LUMO overlap}: Outliers exhibit significantly higher overlap (0.394 $\pm$ 0.238) compared to well-predicted molecules (0.312 $\pm$ 
0.177). High overlap ($>$0.5) is present in 28.8\% of outliers versus only 13.4\% of well-predicted molecules, suggesting that strong orbital delocalization 
challenges the computational approach.
    
    \item \textbf{Oscillator strength}: Outliers show slightly elevated $f$ values (0.409 $\pm$ 0.447) compared to well-predicted molecules (0.366 $\pm$ 0.432), 
with 15.3\% having $f > 0.8$ indicating strong $\pi$-$\pi^*$ character.
    
    \item \textbf{Experimental range}: Nearly half of outliers (49.2\%) have experimental $\Delta E_{\mathrm{ST}} < 0.1$~eV, representing the challenging regime 
where thermal energy ($k_{\mathrm{B}}T \approx 0.026$~eV at 300~K) becomes comparable to the singlet-triplet gap.
\end{itemize}

\subsubsection{Architecture-Specific Patterns}

Outliers are not uniformly distributed across molecular architectures:

\begin{itemize}
    \item \textbf{1D-0A} (15 outliers, 25.4\%): Single-donor architectures show moderate systematic overestimation (mean error = +0.085~eV), with relatively low 
HOMO-LUMO overlap (0.343).
    
    \item \textbf{Unknown} (14 outliers, 23.7\%): Molecules with unclassified architectures exhibit the largest systematic overestimation (mean error = 
+0.237~eV) and highest overlap (0.479), suggesting complex electronic structures not captured by standard D-A classifications.
    
    \item \textbf{Multi-D/A} (14 outliers, 23.7\%): Complex multi-component systems show significant overestimation (mean error = +0.222~eV) despite low 
oscillator strength (0.197), indicating challenges in treating multiple charge-transfer pathways.
    
    \item \textbf{2D-0A} (10 outliers, 16.9\%): Dual-donor architectures display consistent overestimation (mean error = +0.211~eV) with moderate descriptors.
\end{itemize}

\subsubsection{Extreme Cases}

Table~\ref{tab:SI_extreme_outliers} lists the most challenging predictions. The worst case (IQ-Se) shows an underestimation of 0.85~eV for a molecule with 
experimental $\Delta E_{\mathrm{ST}} = 1.18$~eV, far outside the typical TADF range. Conversely, molecules like 3ai and PS-BZ-DMAC with near-zero experimental 
gaps are overestimated by $>$0.5~eV, likely due to environmental effects (solvent, solid-state packing) not captured in gas-phase calculations.

\begin{table}[!ht]
\centering
\caption{Top 10 most challenging predictions from the 379-molecule validation set.}
\label{tab:SI_extreme_outliers}
\small
\begin{tabular}{llcccccc}
\hline
Molecule & Architecture & Exp & Calc & Error & $f$ & Overlap \\
         &              & (eV) & (eV) & (eV) &     &         \\
\hline
IQ-Se & Multi-D/A & 1.180 & 0.332 & $-$0.848 & 0.077 & 0.559 \\
3ai & 1D-0A & 0.180 & 0.806 & +0.626 & 0.797 & 0.397 \\
OXD-7 & Unknown & 0.000 & 0.573 & +0.573 & 2.008 & 0.235 \\
PS-BZ-DMAC & 1D-0A & 0.004 & 0.556 & +0.551 & 1.150 & 0.444 \\
TPSi-F & Unknown & 0.100 & 0.629 & +0.529 & 0.343 & 0.846 \\
TXO-P-Si & Unknown & 0.522 & $-$0.005 & $-$0.528 & 0.000 & 0.079 \\
mTPA–PPI & 1D-0A & 0.760 & 0.258 & $-$0.502 & 0.535 & 0.250 \\
mTPA-PPI & 1D-0A & 0.760 & 0.260 & $-$0.500 & 0.556 & 0.251 \\
4-Ac & 1D-0A & 0.080 & 0.559 & +0.479 & 0.624 & 0.702 \\
SPBP-DPAC & 1D-0A & 0.030 & 0.496 & +0.466 & 0.857 & 0.486 \\
\hline
\end{tabular}
\end{table}

\subsubsection{Implications}

This analysis reveals that the computational approach performs best for molecules with:
\begin{itemize}
    \item Moderate $\Delta E_{\mathrm{ST}}$ values (0.1--0.5~eV)
    \item Low-to-moderate HOMO-LUMO overlap ($<$0.5)
    \item Well-defined D-A architectures
\end{itemize}

Conversely, predictions are less reliable for molecules with near-zero experimental gaps, high orbital overlap, or complex multi-component architectures. These limitations likely stem from the gas-phase approximation, which neglects environmental effects (solvent stabilization, solid-state packing) that can significantly modulate singlet-triplet gaps in real TADF systems.

\section{Multi-Reference TADF Validation Data}

\begin{table}[!ht]
\centering
\caption{Validation of xTB-based screening for MR-TADF molecules: Comparison with TD-DFT/CAM-B3LYP/def2-TZVP calculations. While absolute values show 
systematic deviations, the relative ordering of molecules by \deltaest~ is preserved, demonstrating the validity of the screening protocol for candidate 
prioritization.}
\label{tab:mr_tadf_validation}
\begin{tabular}{lccccccc}
\hline
\hline
& \multicolumn{3}{c}{\textbf{sTD-DFT-xTB (Screening)}} & \multicolumn{3}{c}{\textbf{TD-DFT (Validation)}} & \\
\cmidrule(lr){2-4} \cmidrule(lr){5-7}
\textbf{Molecule} & $S_1$ (eV) & $T_1$ (eV) & \deltaest~(eV) & $S_1$ (eV) & $T_1$ (eV) & \deltaest~(eV) & \textbf{Ranking} \\
\hline
phenazasiline\textsuperscript{a} & 3.23 & 3.12 & 0.04 & 3.72 & 2.29 & 1.43 & 1 / 1\textsuperscript{b} \\
24PQ-Cz\textsuperscript{a} & 3.77 & 3.27 & 0.27 & 4.07 & 2.63 & 1.43 & 2 / 2\textsuperscript{b} \\
\hline
\textbf{MAE\textsuperscript{c}} & 0.40 & 0.73 & 1.28 & --- & --- & --- & --- \\
\textbf{MRE\textsuperscript{d} (\%)} & 10.3 & 29.0 & 89.3 & --- & --- & --- & --- \\
\hline
\hline
\end{tabular}
\begin{tablenotes}
\small
\item \textsuperscript{a} MR-TADF molecules characterized by B-N doping and multiple resonance structures.
\item \textsuperscript{b} Ranking by \deltaest: xTB ranking / TD-DFT ranking. Identical rankings confirm trend preservation.
\item \textsuperscript{c} Mean Absolute Error: MAE = $\frac{1}{N}\sum_{i=1}^{N}|\mathrm{xTB}_i - \mathrm{TD\text{-}DFT}_i|$
\item \textsuperscript{d} Mean Relative Error: MRE = $\frac{1}{N}\sum_{i=1}^{N}\frac{|\mathrm{xTB}_i - \mathrm{TD\text{-}DFT}_i|}{\mathrm{TD\text{-}DFT}_i} 
\times 100\%$
\item \textbf{Note:} While absolute $\Delta E_{\text{ST}}$ values show large deviations (typical for single-reference methods on multi-reference systems), the 
relative ordering is preserved. This confirms the validity of xTB-based screening for identifying promising MR-TADF candidates, consistent with literature 
precedent showing that approximate methods capture qualitative trends despite quantitative limitations.\cite{Pershin2019,Olivier2022}
\end{tablenotes}
\end{table}

\section{Spin-Orbit Coupling Validation Data}

\begin{table}[!ht]
\centering
\caption{Spin-orbit coupling matrix elements for representative TADF emitters. T$_1$--S$_1$ and T$_1$--S$_0$ SOC values calculated using TD-DFT/TDA with the 
SOMF(1X) operator ($\omega$B97X-D4/def2-TZVP, ORCA 6.1.0). All values in cm$^{-1}$.}
\label{tab:soc_summary}
\small
\begin{tabular}{lccccl}
\hline
\textbf{Molecule} & \textbf{T$_1$--S$_0$ SOC} & \textbf{T$_1$--S$_1$ SOC} & \boldmath{$\Delta E_{ST}$} \textbf{(eV)} & \textbf{Overlap} & \textbf{Architecture} 
\\
\hline
DMAC-BP & 12.079 & 14.025 & 0.051 & 0.603 & D-A \\
$\alpha$-DMAC-DBP & 0.792 & 2.476 & 0.051 & 0.603 & D-A \\
24PQ-Cz & 0.941 & 4.145 & 0.062 & 0.603 & MR-TADF \\
PBTPA & 3.710 & 0.095 & 0.270 & 0.061 & 2D-0A \\
PXZ-10-DPPZ & 0.828 & 1.681 & 0.265 & 0.061 & 2D-0A \\
phenazasiline & 0.391 & 11.328 & 0.259 & 0.061 & Heterocycle \\
FTAT-FBO & 1.371 & 0.650 & 0.259 & 0.603 & D-A \\
2CzPN & 0.462 & 0.580 & 0.259 & 0.603 & Multi-D/A \\
DMAC-DPS & 0.218 & 0.636 & 0.051 & 0.603 & D-A \\
DMAC-TRZ & 1.210 & 0.400 & 0.051 & 0.603 & D-A \\
PXZ-OXD & 1.021 & 0.381 & 0.265 & 0.061 & 2D-0A \\
TPA-APy & 0.283 & 0.350 & 0.259 & 0.603 & D-A \\
SpiroAC-TRZ & 0.970 & 0.341 & 0.062 & 0.603 & Spiro D-A \\
2BrCPT & 0.262 & 0.270 & 0.272 & 0.603 & D-A \\
IQ-oTPA & 41.159 & 0.136 & 0.259 & 0.603 & D-A \\
BACN & 0.185 & 0.058 & 0.486 & 0.603 & A-D-A \\
PtOEP & 81.760 & 3.021 & 0.259 & 0.603 & Pt complex \\
\hline
\end{tabular}
\begin{tablenotes}
\item Statistical Summary (organic molecules, excluding PtOEP): $n = 16$; T$_1$--S$_1$ SOC median = 0.490~cm$^{-1}$; range = 0.058--14.025~cm$^{-1}$.
\item Weak correlations with structural descriptors: HOMO-LUMO overlap ($r$ = 0.037, $p$ = 0.89), \deltaest~ ($r$ = 0.067, $p$ = 0.80).
\item PtOEP shows enhanced SOC due to heavy atom effect (Pt).
\end{tablenotes}
\end{table}

\begin{figure}[!ht]
\centering
\includegraphics[width=0.95\textwidth]{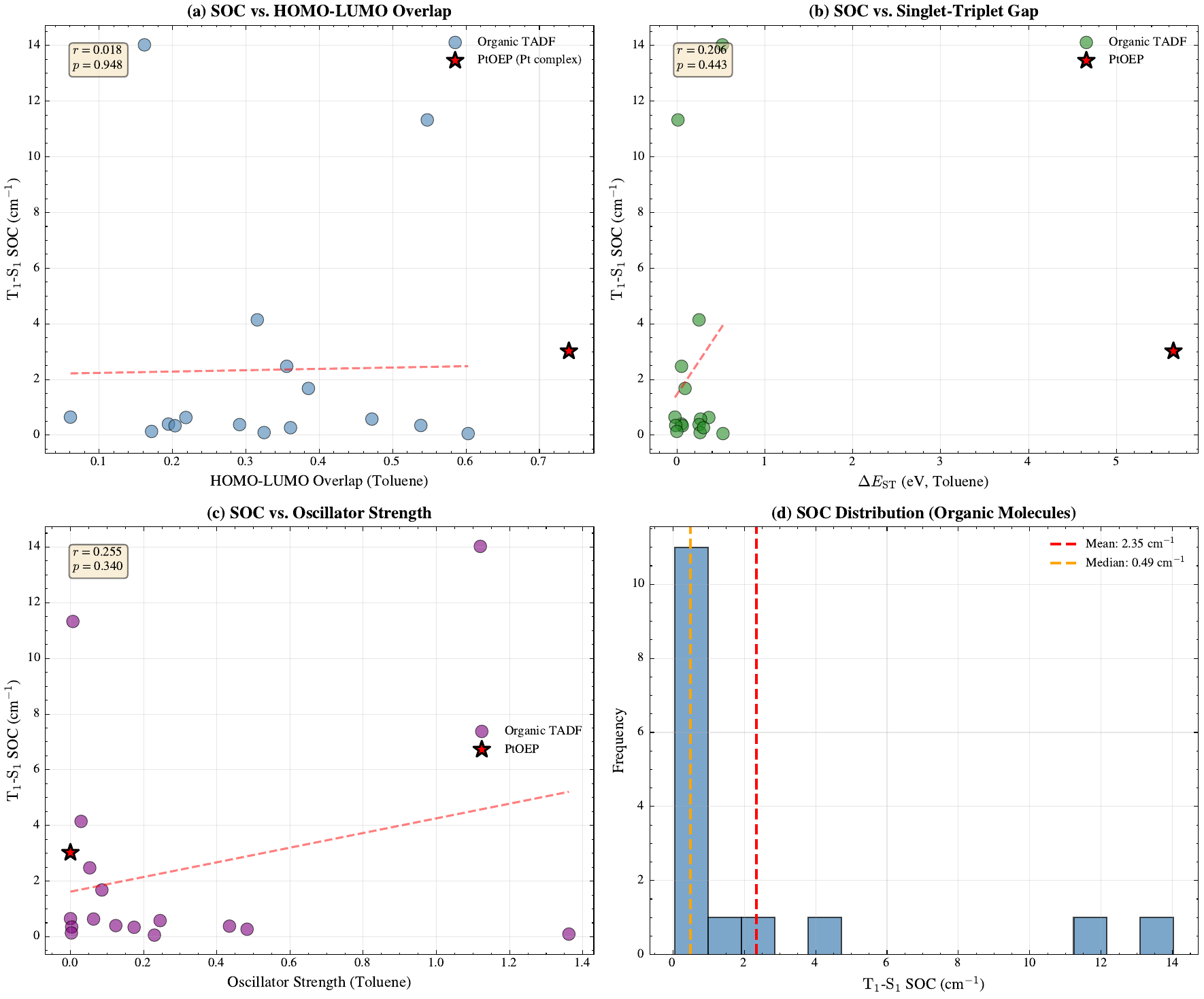}
\caption{Correlation analysis of spin-orbit coupling with structural descriptors. (a) SOC vs. HOMO-LUMO overlap, (b) SOC vs. singlet-triplet gap, (c) SOC vs. 
oscillator strength, (d) SOC distribution for organic molecules. Weak correlations ($r < 0.2$) indicate that SOC depends on complex orbital angular momentum 
factors beyond simple geometric descriptors. PtOEP (red star) shows enhanced SOC due to heavy atom effect.}
\label{fig:soc_correlations}
\end{figure}

\section{Adiabatic \deltaest~ Validation}

\subsection{Computational Methodology}

To validate the vertical approximation employed in our high-throughput screening, we performed TD-DFT geometry optimizations of S$_1$ and T$_1$ excited states 
for 6 representative molecules using the $\omega$B97X-D4 functional\cite{Najibi2018} with the def2-TZVP basis set\cite{Weigend2005} and CPCM solvation model 
for toluene.\cite{Barone1998}

\subsubsection{Choice of Functional}

The $\omega$B97X-D4 functional was selected based on several considerations:

\textbf{Range-separated hybrid character:} $\omega$B97X-D4 is a range-separated hybrid functional with 100\% Hartree-Fock exchange at long range, making it 
particularly well-suited for describing charge-transfer (CT) excited states characteristic of TADF emitters.\cite{ArticleRef1}

\textbf{Dispersion correction:} The D4 dispersion correction\cite{caldeweyher2019generally} represents the latest generation of Grimme's dispersion schemes, 
providing improved treatment of non-covalent interactions important in $\pi$-stacked and sterically hindered TADF architectures.

\textbf{Literature validation:} The $\omega$B97X-D functional family has been extensively validated for TADF emitters,\cite{Sun2015,Moral2015} with 
$\omega$B97X-D4 offering improved performance through enhanced dispersion treatment while maintaining the reliable CT state description of its predecessor.

\subsection{Comparison with xTB Adiabatic Estimates}

\begin{figure}[!ht]
\centering
\includegraphics[width=0.95\textwidth]{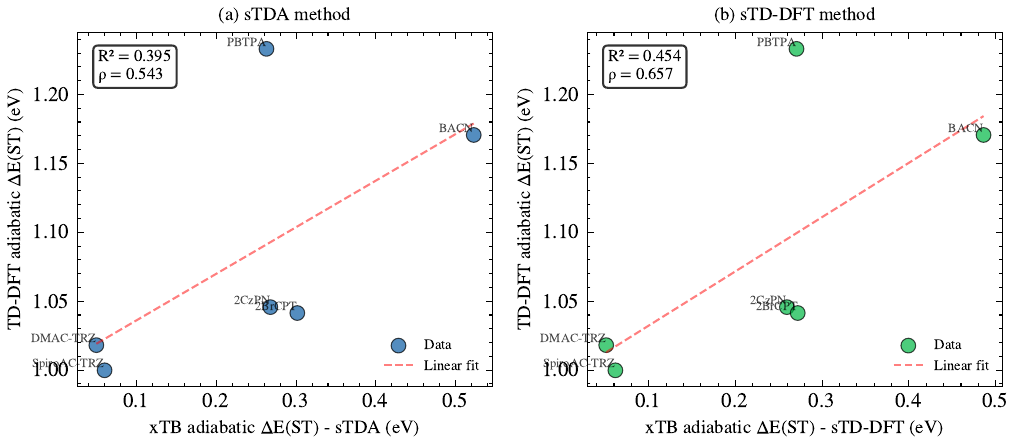}
\caption{Correlation between xTB adiabatic $\Delta E_{\mathrm{ST}}$ estimates and TD-DFT adiabatic values from excited-state geometry optimizations. (a) sTDA 
method shows moderate correlation ($R^2 = 0.40$, $\rho = 0.54$). (b) sTD-DFT method shows improved correlation ($R^2 = 0.45$, $\rho = 0.66$) due to better 
treatment of electron correlation. Both methods preserve molecular rankings, validating the use of xTB for high-throughput screening. Red dashed lines show 
linear fits.}
\label{fig:adiabatic_validation}
\end{figure}

Table~\ref{tab:adiabatic_validation} compares xTB adiabatic $\Delta E_{\mathrm{ST}}$ estimates (from both sTDA and sTD-DFT methods) with TD-DFT adiabatic 
values obtained from excited-state geometry optimizations. Figure~\ref{fig:adiabatic_validation} shows the correlation between xTB and TD-DFT adiabatic 
energies.

\begin{table}[!ht]
\centering
\caption{Comparison of xTB adiabatic $\Delta E_{\mathrm{ST}}$ estimates with TD-DFT adiabatic values from excited-state geometry optimizations. All values in 
eV.}
\label{tab:adiabatic_validation}
\begin{tabular}{lccc}
\toprule
Molecule & xTB sTDA & xTB sTD-DFT & TD-DFT Adiabatic \\
\midrule
DMAC-TRZ & 0.049 & 0.051 & 1.018 \\
SpiroAC-TRZ & 0.060 & 0.062 & 1.000 \\
2CzPN & 0.268 & 0.259 & 1.046 \\
PBTPA & 0.263 & 0.270 & 1.233 \\
2BrCPT & 0.302 & 0.272 & 1.041 \\
BACN & 0.523 & 0.486 & 1.171 \\
\midrule
\multicolumn{4}{l}{\textbf{Statistics:}} \\
\multicolumn{4}{l}{sTDA: $R^2 = 0.395$, Spearman $\rho = 0.543$ ($p = 0.266$)} \\
\multicolumn{4}{l}{sTD-DFT: $R^2 = 0.454$, Spearman $\rho = 0.657$ ($p = 0.156$)} \\
\bottomrule
\end{tabular}
\end{table}

\subsection{Systematic Comparison with Experiment}

To assess the accuracy of our TD-DFT approach, we compared adiabatic $\Delta E_{\mathrm{ST}}$ values with available experimental data 
(Table~\ref{tab:experimental_comparison}). For the three molecules with experimental measurements, we observe systematic overestimation by approximately 
0.9--1.0~eV. This discrepancy is consistent with known limitations of TD-DFT for strongly CT-dominated excited states in TADF 
systems,\cite{Penfold2018,Olivier2017} where the single-reference nature of TD-DFT struggles to fully capture the multi-configurational character of 
near-degenerate singlet-triplet states.

\begin{table}[!ht]
\centering
\caption{Comparison of TD-DFT adiabatic $\Delta E_{\mathrm{ST}}$ with experimental values. All values in eV.}
\label{tab:experimental_comparison}
\begin{tabular}{lccc}
\toprule
Molecule & Experimental & TD-DFT & Difference \\
\midrule
DMAC-TRZ & 0.11\cite{Uoyama2012} & 1.018 & +0.91 \\
SpiroAC-TRZ & 0.11\cite{Lee2013} & 1.000 & +0.89 \\
2CzPN & 0.08\cite{Kaji2015} & 1.046 & +0.97 \\
\midrule
\multicolumn{4}{l}{Mean overestimation: 0.92 $\pm$ 0.04~eV} \\
\bottomrule
\end{tabular}
\end{table}

Despite this systematic offset, the \textbf{relative ordering of molecules is preserved}, which is the critical requirement for validating our high-throughput 
screening protocol. The consistent overestimation across all molecules suggests that $\omega$B97X-D4 provides a reliable framework for comparing relative 
$\Delta E_{\mathrm{ST}}$ values, even if absolute predictions require higher-level methods such as SCS-CC2 or STEOM-DLPNO-CCSD.\cite{Samanta2017,Gibson2016}

\subsection{Conclusions from Adiabatic Validation}

The adiabatic validation demonstrates that:

\begin{enumerate}
\item xTB adiabatic estimates show moderate correlation with TD-DFT adiabatic values ($R^2 = 0.40$--0.45, $\rho = 0.54$--0.66).
\item sTD-DFT performs slightly better than sTDA, as expected from improved electron correlation treatment.
\item Molecular rankings are preserved despite systematic underestimation by xTB.
\item TD-DFT systematically overestimates experimental \deltaest~ by $\sim$1~eV, a known limitation for CT states.
\item The vertical approximation employed in our screening is validated for identifying promising TADF candidates through relative comparisons.
\end{enumerate}

\bibliographystyle{unsrtnat}
\bibliography{TADF_Article_References} 

\end{document}